\documentclass[aps,amsmath,pre,superscriptaddress,showpacs]{revtex4}

\usepackage{graphicx}
\usepackage{epsfig}
\usepackage{amsthm, amsmath, amsfonts, ae}

\begin{document}

\title{A hard-sphere model on generalized Bethe lattices: Statics}
 
\author{Hendrik Hansen-Goos} 
\affiliation{Institut f\"ur Theoretische Physik, Universit\"at 
G\"ottingen, Friedrich-Hund-Platz 1, D-37077 G\"ottingen, Germany} 
\author{Martin Weigt} 
\affiliation{Institute for Scientific Interchange, Viale Settimio 
Severo 65, I-10133 Torino, Italy}

\date{\today}

\begin{abstract}
We analyze the phase diagram of a model of hard spheres of chemical
radius one, which is defined over a generalized Bethe lattice
containing short loops. We find a liquid, two different crystalline, a
glassy and an unusual crystalline glassy phase. Special attention is
also paid to the close-packing limit in the glassy phase. All
analytical results are cross-checked by numerical Monte-Carlo
simulations.
\end{abstract}
\pacs{64.70.Pf,64.60.Cn,75.10.Nr}

\maketitle

\section{Introduction}

Even after many years of vivid interest, the structural glass
transition is still an open and alive topic of research
\cite{Go,An,BoCuKuMe,RiSo,Cu}. Being defined by a drastic slowing
down of the equilibration time of a liquid, it is introduced as a
dynamical, non-equilibrium phenomenon. Up to now it is, however, one
of the crucial questions if this slowing down is accompanied by an
equilibrium glass transition or not.

Recently, lattice-gas models have played an increasing role in this
discussion \cite{biroli,weigt,pica,rivoire}. The central idea
of these models is to incorporate {\it static geometrical constraints}
via hard-sphere interactions which restrict the possible particle
packings and introduce some kind of geometrical frustration. Defined
on (generalized) Bethe lattices, these models are characterized by the
existence of a dynamical glass transition, followed by a static one.

These models are contrasted by the so-called {\it kinetically
constrained} models \cite{RiSo}, in which not all particle moves are
permitted, but which are characterized by a trivial thermodynamic
equilibrium behavior. Defined on a Bethe lattice, they show, however,
a very similar dynamical behavior to the above mentioned models
\cite{SeBiTo}.

The model analyzed in this paper was introduced in \cite{weigt}, and
it belongs to the class of geometrically constrained models. Its
equilibrium behavior will be analyzed in great detail using the cavity
method \cite{mezparvir,mezpar}. The model can be defined over any
lattice or graph. Sites are either occupied by particles, or they are
empty. The particles are, however, too large to allow any two
neighboring sites to be occupied simultaneously. In the presence of
short loops in the lattice, this hard-core exclusion is sufficient to
create a very rich phase diagram, including a liquid, two different
crystalline, a glassy and an unusual, crystalline glassy
phase. Whereas this paper concentrates completely on the static
properties of the model, a subsequent publication \cite{self} will
consider the dynamics.

The paper is organized as follows: In the next section, the model is
defined. Sec.~III is dedicated to a description of an iterative method
for the calculation of the partition function, or more precisely of
effective local fields. Sec.~IV solves this iterative method for the
liquid-crystal transition, whereas Secs.~V and VI are dedicated to
the glass phase. Conclusion and outlook are given in the last section.

\section{The model}
\label{sec:model}

The model was already introduced and studied in \cite{weigt}. In its
most general formulation, it is defined on a graph $G$ having
vertices $V=\{1,...,N\}$ and undirected edges $\{i,j\} \in E$
connecting pairs of vertices.  Vertices can either be empty, $n_i=0$,
or they are occupied by a particle, $n_i=1$. These particles interact
via a hard core of chemical radius one, i.e. neighboring vertices
cannot be occupied simultaneously by two particles. Formulated in a
more formal way, all edges $\{i,j\}\in E$ fulfill the constraint $n_i
n_j = 0$, i.e. one of the end vertices has to be empty. The model can
thus be characterized by its grand-canonical partition function
\begin{equation}
\label{eq:partition_function}
  \Xi(\mu) = \sum_{n_1, \ldots , n_N \in \{0, 1\}}
  e^{\mu\sum_{i=1}^{N}n_i}\,\prod_{\{i,j\}\in E} (1-n_i n_j) \, ,
\end{equation}  
or its grand-canonical potential
\begin{equation}
\label{eq:Omega}
  \Omega = -\frac 1\mu \ln \Xi \, .
\end{equation}  
Here we have introduced the chemical potential $\mu$ which is coupled
to the total particle number $\sum_i n_i$ and can thus be used to
regulate the particle density
\begin{equation}
\label{eq:density}
\rho = \frac 1N \sum_{i=1}^{N}n_i \, .
\end{equation}  
Note that the product over all edges in
Eq.~(\ref{eq:partition_function}) serves as an indicator function for
allowed particle configurations: Whenever there is at least one pair
of occupied adjacent vertices, the product vanishes and thus the
configuration does not contribute to the partition function. Note also
that due to the hard-core interaction between particles the
temperature does not play any role in the model, without loss of
generality it is set to one.

\begin{figure}[tbp]
  \begin{center}
    \includegraphics[height = 8cm]{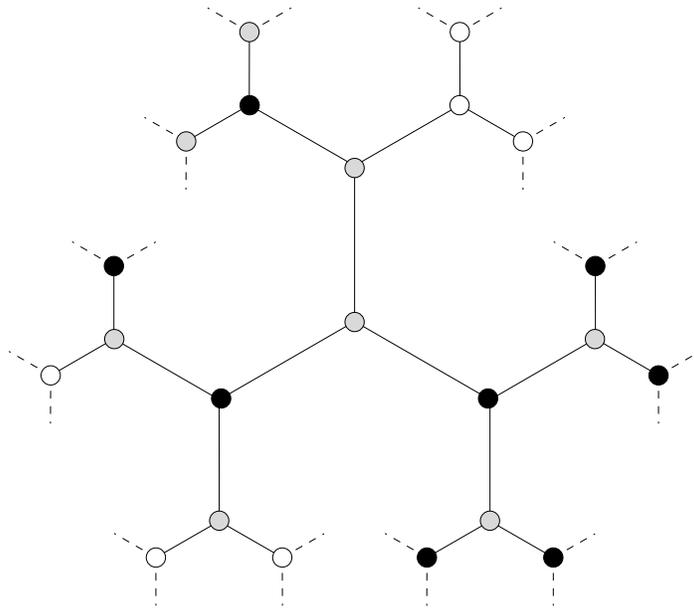}
    \caption{Part of a Bethe lattice with $k=2$. Black vertices mark
    occupied sites, gray ones cannot be occupied due to volume
    exclusion. The lower right branch shows the densest local
    packing.}
\label{fig_bethegitter}
  \end{center}
\end{figure}

So far, the model is still defined on a general graph. In this work we
will concentrate on (generalized) Bethe lattices. Bethe lattices can
be defined in two different ways:
\begin{itemize}
\item The first definition considers a Bethe lattice as an infinite
regular tree, i.e. a cycle-free graph of fixed vertex degree (or
coordination number) $k+1$. The hard-sphere model on this graph was
first considered by Runnels \cite{runnels}.
\item As a second possibility we can define a Bethe lattice as a
random $(k+1)$-regular graph of $N$ vertices, with $N\to\infty$ in the
thermodynamic limit. This version is frequently used in the context of
glassy systems \cite{mezpar}, and allows in particular for the
simulation of finite systems.
\end{itemize}
Locally, both systems are equivalent due to the fixed vertex degree,
the case $k=2$ together with a possible particle packing is visualized
in Fig. \ref{fig_bethegitter}. The crucial difference results from the
existence of cycles in random graphs: Even if these are of length
${\cal O}(\ln N)$, i.e. their length diverges for large $N$,
they are in general inconsistent with the crystalline structure which
will be considered in Sec. \ref{sec_glstatik}. A densest particle
packing on a Bethe lattice defined as an infinite tree consists of an
alternation of occupied and empty sites. In random regular graphs,
many cycles of odd length exist, preventing this alternating
configuration from being globally feasible. A way out of this dilemma
will be discussed later on.

\begin{figure}[tbp]
  \begin{center}
    \includegraphics[height = 9cm]{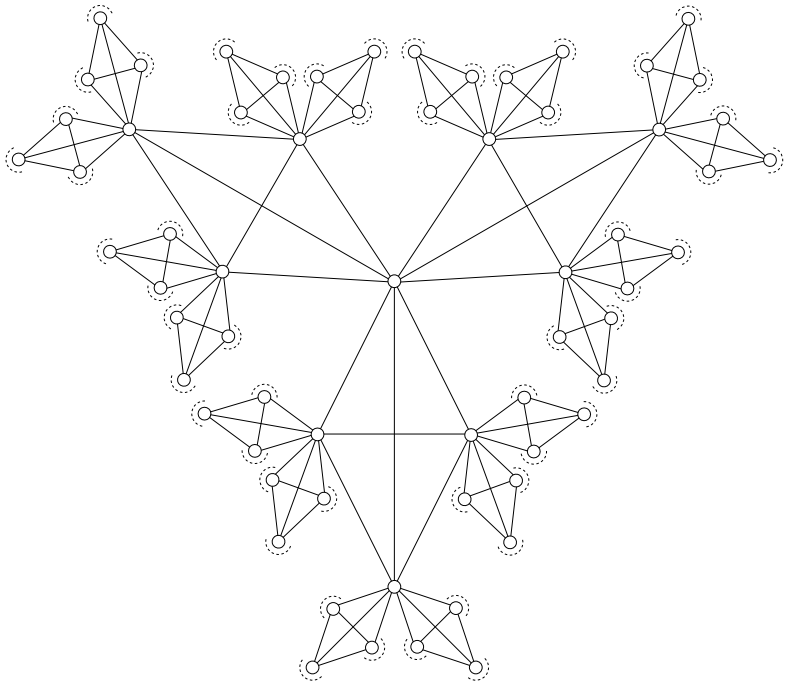}
    \caption{Part of a generalized Bethe-lattice with $k=2$,
    $p=3$. The 21 cliques connecting the central vertex to it's
    nearest and second neighbors are shown.}
\label{fig_allgbethe}
  \end{center}
\end{figure}

Realistic, i.e. finite-dimensional, systems contain, however, many
short cycles. To imitate this, we generalize Bethe lattices in the
following way, cf. also Fig.~\ref{fig_allgbethe}: The graph is
composed of cliques, i.e. fully connected subgraphs of $p+1$ vertices
each. In each vertex, $k+1$ of these cliques merge, such that the
resulting graph has constant vertex degree $p(k+1)$. The global
structure resembles, however, a Bethe lattice: We assume that besides
the loops inside the cliques there are no other short cycles. Note
that ordinary Bethe lattices are obtained in the special case $p=1$,
which is included in the following discussion of the general case.
Also generalized Bethe lattices can be defined in the two ways
discussed above, with similar consequences on the global feasibility
of locally dense packings.

\section{Iteration of the partition function}

\begin{figure}[tbp]
  \begin{center}
    \includegraphics[height = 6cm]{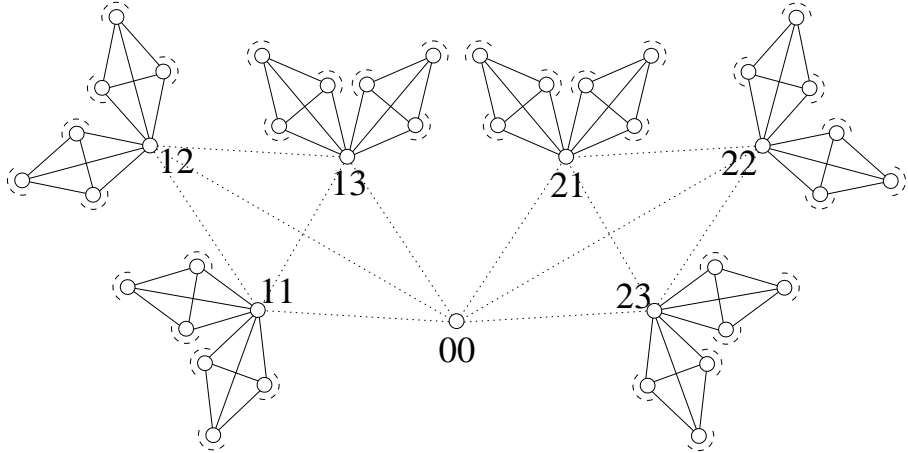}
    \caption{Iteration of rooted trees: $k=2$ branches of $p=3$ rooted
      trees are merged to form a new tree with root $00$.}
\label{fig_betherek}
  \end{center}
\end{figure}

The direct calculation of the partition function is complicated. It can,
however, be achieved via an iterative method. For doing so, we shall
introduce the notation of a {\it rooted subtree}: Imagine one of the
cliques containing a vertex $i$ to be removed from the graph, then a
rooted subtree consists of the connected component containing $i$. The
vertex $i$ itself is denoted as the root of the subtree,
cf. Fig.~\ref{fig_betherek}. Note that the root has only $kp$
neighbors, whereas all other vertices in the subtree have degree
$(k+1)p$. For each of these rooted trees we introduce two restricted
partition functions: $\Xi_e^{i}$ denotes the partition function with
$n_i$ fixed to zero, whereas $\Xi_*^{i}$ corresponds to an occupied
root.

Given $k p$ rooted trees ($j=1, \ldots , k$, $l = 1, \ldots , p$),
these can be composed according to Fig.~\ref{fig_betherek} to form
one new rooted tree with root $j=0$, $l=0$. The double index is
introduced to lighten the notation in the following equations: The
first index defines the (removed) clique containing the root, whereas
$k$ enumerates the vertices within this clique. The restricted
partition function of the new tree can now be easily inferred from
those of the $kp$ original ones:
\begin{subequations}
\label{gl_rekzsumme}
\renewcommand{\theequation}{\theparentequation \roman{equation}}
\begin{eqnarray}
\Xi_e^{00} & = & \prod_{j=1}^{k} \Biggl( \prod_{l=1}^{p}\Xi_e^{jl} +
  \sum_{l=1}^{p} \Xi_*^{jl} 
      \prod_{\substack{m=1\\m \neq l}}^{p}\Xi_e^{jm} \Biggr) \\
\Xi_*^{00} & = & e^{\mu}\, \prod_{j=1}^{k} 
      \Biggl(\prod_{l=1}^{p}\Xi_e^{jl}\Biggr) \, .
\end{eqnarray}
\end{subequations}
The expression in the brackets stands for the contribution of one
branch: In case that the root $00$ is empty, up to one of the
neighbors in each branch can be occupied. If, on the other hand, the
root is occupied, none of its neighbors is allowed to be occupied. The
additional factor $e^\mu$ in the second equation takes into account
the chemical potential acting on the root particle.

Introducing now the {\it cavity fields}
\begin{equation} \label{gl_cavfieldx}
  x^{jl} \doteq \frac{\Xi_*^{jl}}{\Xi_e^{jl}}
\end{equation}
(more precisely, $\ln (x^{jl}+1)/\mu$ defines a local field coupled to 
$n_{jl}$), we find
\begin{eqnarray}\label{gl_lokfeldrek}
  x^{00} & = & \frac{e^{\mu}}
           {\prod_{j=1}^{k}\bigl( 1 + \sum_{l=1}^{p} x^{jl}\bigr)}
           \nonumber\\
         & \doteq & \hat{x}(x^{11}, x^{12}, \ldots , x^{kp}) .
\end{eqnarray}
In order to understand the thermodynamic behavior of our model, we
have to find appropriate solutions to these equations. In
the next section, we shall discuss the case of the transition toward
an ordered, crystalline phase, whereas Sec.~\ref{chap_1RSB} is
dedicated to glassy solutions.

Before doing so, we have to clarify the physical meaning of
$x^{jl}$ and its connection to observables of the system. A simple
observable is given by the local density $\rho^{00}$ of vertex
$00$. This quantity can be obtained by composing rooted subtrees as well.
In the original graph, each vertex has, however, $(k+1)p$
neighbors. Within one iteration step, this can be achieved by merging $(k+1)$
instead of $k$ branches. Denoting the corresponding partition
functions by $\tilde{\Xi}_{e/*}^{00}$, we get
\begin{subequations}
\renewcommand{\theequation}{\theparentequation \roman{equation}}
\begin{eqnarray}
\tilde{\Xi}_e^{00} & = & \prod_{j=1}^{k+1} \Biggl( \prod_{l=1}^{p}\Xi_e^{jl} +
  \sum_{l=1}^{p} \Xi_*^{jl} 
      \prod_{\substack{m=1\\m \neq l}}^{p}\Xi_e^{jm} \Biggr) \\
\tilde{\Xi}_*^{00} & = & e^{\mu}\, \prod_{j=1}^{k+1} 
      \Biggl(\prod_{l=1}^{p}\Xi_e^{jl}\Biggr) \, .
\end{eqnarray}
\end{subequations}

This results in
\begin{equation}\label{gl_connlfobs}
  \frac{\rho^{00}}{1-\rho^{00}} =
  \frac{\tilde{\Xi}_*^{00}}{\tilde{\Xi}_e^{00}} = \frac{e^{\mu}}
    {\prod_{j=1}^{k+1}\bigl( 1 + \sum_{l=1}^{p} x^{jl}\bigr)} \, ,
\end{equation}
i.e. physical observables can be directly calculated from the cavity
fields.

\section{Crystallization}
\label{sec_glstatik}

\subsection{Liquid and crystalline solutions}

To solve Eq.~(\ref{gl_lokfeldrek}), we have to restrict the solution
space by considering the limiting cases $\mu\to\pm\infty$. For the
moment we discuss only the first case for the definition of the
generalized Bethe lattice, i.e. we exclude the existence of long loops
which may have some influence on the global packing structure.

The limiting case of an empty (or very dilute) system ($\mu \to
-\infty$) is characterized by a spatially homogeneous density
describing a {\it liquid} phase. Consequently, the cavity fields are all equal
to some $x^*$ given self-consistently by $x^* = \hat{x}(x^*, \ldots ,
x^*)$. 

In the limit $\mu\to\infty$ we have to search for close-packings of
the system. In these every clique carries exactly one particle, due to
their regular structure these configurations are to be considered as
being {\it crystalline}. The number of these configurations is
exponential in the size of the graph ({\em i.e.} the number of sites), as can be seen easily: Initializing the configuration
by putting one particle on an arbitrary vertex, all neighboring
vertices have to be free. The $(k+1)kp^2$ second neighbors are
partitioned into $(k+1)kp$ cliques of $p$ vertices each, and each of
these cliques can carry exactly one particle - so there are
$p^{(k+1)kp}$ possible configurations for selecting the particle
positions in between the second neighbors. Iterating this argument, we
obviously find an exponential number of close-packings.

\begin{figure}[tbp]
  \begin{center}
    \includegraphics[height = 3.7cm]{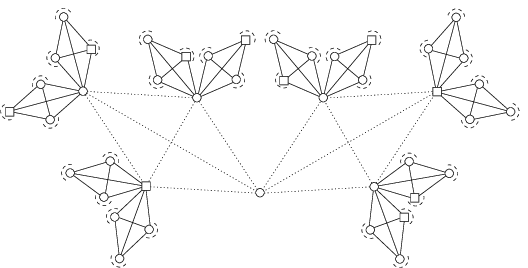}
    \hspace{0.4cm}
    \includegraphics[height = 3.7cm]{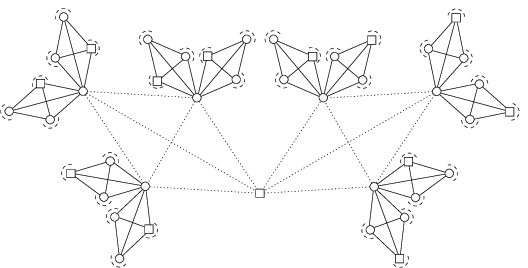}
    \caption{Possible iteration steps leading to rooted trees with
      root in the 0-lattice (left), 1-lattice (right). Vertices
      belonging to the 0-lattice (1-lattice) are depicted by $\circ$ ($\Box$).}
\label{fig_ocemiter}
  \end{center}
\end{figure}

Selecting one close-packing, we can identify two sub-lattices: The
first is formed by the occupied vertices (1-lattice), the second by
the empty vertices (0-lattice). We introduce two cavity fields
$x^{(0)}$ and $x^{(1)}$ for the sub-lattices. Note that a more general
ansatz that works with $p+1$ different cavity fields corresponding one
to each site of a clique might lead to more complicated scenarios on
$(p+1)$-partite graphs. It can be shown, however, that the cavity fields
can only take two different values. We conclude that at least for the
case $p \le 2$ no extra solution appears.

For the iteration of $x^{(0)}$ and $x^{(1)}$ we
have to distinguish two cases, represented in Fig.~\ref{fig_ocemiter}:
Either the root belongs to the 1-lattice, and has only neighbors from
the 0-lattice, or it belongs to the 0-lattice, and has exactly one
1-neighbor and $p-1$ 0-neighbors in each clique. The resulting
equations thus read:
\begin{subequations}
\label{gl_lfrekallg}
\renewcommand{\theequation}{\theparentequation \roman{equation}}
\begin{align}
  x^{(0)} & = \hat{x}(x^{(1)}, \underbrace{x^{(0)}, \ldots ,
    x^{(0)}}_{p-1}, \ldots) \\
  x^{(1)} & = \hat{x}(\underbrace{x^{(0)}, \ldots , x^{(0)}}_{p}, \ldots),
\end{align}
\end{subequations}
where the presented block of variables has to be repeated $k$ times in
the argument of $\hat x$. The permutation symmetry of variables inside
one block is already used here. The equations can be rewritten
explicitly as
\begin{subequations}
\label{gl_lfrekcry}
\renewcommand{\theequation}{\theparentequation \roman{equation}}
\begin{align}
  x^{(0)} & = \frac{e^{\mu}}
  {\bigl( 1 + x^{(1)} + (p-1)\,x^{(0)}\bigr)^k} \\
  x^{(1)} & = \frac{e^{\mu}}
  {\bigl( 1 + p\,x^{(0)}\bigr)^k} \,.
\end{align}
\end{subequations}
Using Eq.~\ref{gl_connlfobs} we thus find to sub-lattice densities 
\begin{subequations}
\label{gl_connlfobscry}
\renewcommand{\theequation}{\theparentequation \roman{equation}}
\begin{align}
  \frac{\rho^{(0)}}{1-\rho^{(0)}} & = \frac{e^{\mu}}
  {\bigl( 1 + x^{(1)} + (p-1)\,x^{(0)}\bigr)^{k+1}} \\
  \frac{\rho^{(1)}}{1-\rho^{(1)}} & = \frac{e^{\mu}}
  {\bigl( 1 + p\,x^{(0)}\bigr)^{k+1}} \, .
\end{align}
\end{subequations}
Eliminating the cavity fields using Eq.~(\ref{gl_lfrekcry}), we find the
{\it equilibrium equations of state} of our model:
\begin{subequations}
\label{gl_lsgstatik}
\renewcommand{\theequation}{\theparentequation \roman{equation}}
\begin{align}
  e^{\mu} & = \frac{\rho^{(0)} \bigl(1-\rho^{(0)}\bigr)^k}
  {\bigl(1-\rho^{(1)}-p\,\rho^{(0)}\bigr)^{k+1}} \\
  e^{\mu} & = \frac{\rho^{(1)} \bigl(1-\rho^{(1)}\bigr)^k}
  {\bigl(1-\rho^{(1)}-p\,\rho^{(0)}\bigr)^{k+1}}\,.
\end{align}
\end{subequations}
For fixed $k$ and $p$, they imply functions $\rho^{(0)}(\mu)$ and
$\rho^{(1)}(\mu)$. Together with the grand-canonical potential which
will be calculated in the following Sec.~\ref{sec_bethepeierls}, they
allow to determine the equilibrium behavior and, in particular,
location and order of the crystallization transition.

Let us therefore discuss the solutions of Eqs.~\ref{gl_lsgstatik},
coming first back to the liquid solution. It can be understood as the
special case of equal sub-lattice densities,
$\rho^{(0)}=\rho^{(1)}=\rho$. We thus have to solve just one global
equation of state,
\begin{equation}\label{gl_lsgstatikfl}
  e^{\mu} = \frac{\rho \bigl(1-\rho\bigr)^k}
  {\bigl(1-(p+1)\,\rho\bigr)^{k+1}} \, .
\end{equation}
It has just one solution, growing monotonously from $\rho=0$ for $\mu
\to -\infty$ to the maximally possible density $\rho=\frac{1}{p+1}$
for $\mu \to \infty$.

Concentrating now on crystalline solutions with
$\rho^{(0)}\neq\rho^{(1)}$, we find that Eqs.~(\ref{gl_lsgstatik})
lead to the $p$-independent equilibrium condition
\begin{equation}\label{gl_gleichgewbed}
  \rho^{(0)}\,(1-\rho^{(0)})^k = \rho^{(1)}\,(1-\rho^{(1)})^k.
\end{equation}
The function $r(1-r)^k$ has a single maximum in $r=\frac1{k+1}$, thus
for any $\rho^{(1)}\neq\frac1{k+1}$ there is a solution for
$\rho^{(0)}$ being different from $\rho^{(1)}$, only for
$\rho^{(1)}=\frac1{k+1}$ both sub-lattice densities have to coincide.
This defines a function
$\rho^{(0)}=\phi(\rho^{(1)})$. Eqs.~\ref{gl_lsgstatik} can thus be reduced
to the single equation
\begin{equation*}
  e^{\mu} = \frac{\rho^{(1)}\,(1-\rho^{(1)})^k}
    {(1-\rho^{(1)}-p\,\phi(\rho^{(1)}))^{k+1}} \, .
\end{equation*} 
In order to find the spinodal point $\mu_{sp}$, i.e.~the minimal value
of $\mu$ where the last equation has a solution, we have to minimize
the right-hand side of this equation with respect to $\rho^{(1)}$.

For $\rho^{(1)}\ne\frac{1}{k+1}$ this leads to a condition on the
sub-lattice densities exactly at the spinodal point,
\begin{equation}\label{gl_spinobed}
 1-\rho_{sp}^{(1)}-p\,\rho_{sp}^{(0)} -\rho_{sp}^{(0)} (k+1)
 \bigl((kp-1)\rho_{sp}^{(1)}-p+1\bigr)\,\overset{!}{=}\,0 \, ,
\end{equation}
which is to be completed by the equilibrium condition $\rho_{sp}^{(0)}
= \phi(\rho_{sp}^{(1)})$. Taking as an example the case $k=2$, $p=2$,
there is exactly one solution to these equations: $\rho^{(0)}_{sp} =
\frac{1}{9}(6-\sqrt{33}) \simeq 0.028$ and $\rho^{(1)}_{sp} =
\frac{1}{18}(9+\sqrt{33}) \simeq 0.819$. The global density results in
$\rho_{sp} = (p\rho^{(0)}_{sp}+\rho^{(1)}_{sp})/(p+1) \simeq
0.292$. From Eqs.~(\ref{gl_lsgstatik}) we infer the chemical potential
$\mu_{sp} \simeq 2.64$ of the spinodal point. Plugging this into
Eq.~(\ref{gl_lsgstatikfl}) in order to determine the liquid density at
the same chemical potential, we find $\rho_{\text{liquid}}(\mu_{sp}) \simeq
0.261$. This value is substantially smaller than $\rho_{sp}$, we thus
conclude that the crystalline solution appears {\it discontinuously}
at $\mu_{sp}$. Note also that, for $\mu>\mu_{sp}$ two different
solutions for the sub-lattice densities are evolving, one with a
monotonously increasing global density, the other one with an
initially decreasing global density.

For $\rho_{sp}^{(1)} = \frac{1}{k+1}$, on the other hand, we know that
$\rho_{sp}^{(0)} = \phi(\rho^{(1)}_{sp}) = \frac{1}{k+1}$ coincides
with $\rho_{sp}^{(1)}$. In this case, the crystalline solution appears
{\it continuously} out of the liquid
solution. Eqs.~(\ref{gl_lsgstatik}) are, however, consistent with this
solution if and only if $p=1$. The continuous {\em onset} of a
crystalline solution can therefore be found only on usual
Bethe-lattices. This statement does not include the scenario where an
unstable crystalline solution becomes stable at $\rho^{(0)} =
\rho^{(1)} = \frac{1}{k+1}$ which will be discussed later on.

\subsection{The grand-canonical potential}
\label{sec_bethepeierls}

In order to find out which of the solutions discussed above is the
thermodynamically stable one, and if there are further metastable
phases, we have to compare the corresponding values of the
grand-canonical potential, or more precisely its density $\omega$ per
clique. Due to the distinction of the two types of vertices, we can
write this as $\omega = p \omega^{(0)} + \omega^{(1)}$, where
$\omega^{(0/1)}$ are the potentials per 0/1-site. Knowing the cavity
fields $x^{(0)}$ and $x^{(1)}$, $\omega$ can be calculated using the
following construction:
\begin{itemize}
\item We start with $p(k+1)$ generalized Bethe-lattices.
\item We choose one clique in each lattice, and remove all its edges.
\item We obtain $(p+1) \cdot p(k+1)$ rooted trees, among which $p
  \cdot p(k+1)$ have a 0-root, and $p(k+1)$ have a 1-root.
\item We add $p$ 0-vertices, and connect each one with $k+1$ of the
  1-trees and $(p-1)(k+1)$ of the 0-trees, such that we obtain $p$ new
  generalized Bethe-lattices.
\item We add one 1-vertex, and connect it to the remaining $p(k+1)$ 
  0-trees, such that we get one more generalized Bethe-lattice.
\item In this way, we obtain $p+1$ lattices.
\end{itemize}
We have to sum up the changes in the grand-canonical potential induced
by following the construction: We have added $p$ 0-vertices and one
1-vertex. To do so, we had to remove the edges of $p(k+1)$ cliques;
denote the change in the potential induced by adding the edges of a
clique by $\Delta\Omega_{\text{link}}$. Further on we had to connect
the newly added vertices and the corresponding rooted trees; the
change in potential induced by one of these steps is denoted by
$\Delta\Omega^{(0)}_{\text{site}}$
resp.~$\Delta\Omega^{(1)}_{\text{site}}$ according to the added
vertex. The total bilance thus reads:
\begin{equation} \label{gl_bethepeierlsapprox}
 \omega = p\omega^{(0)}+\omega^{(1)} = -p(k+1)\Delta\Omega_{\text{link}}
                           + p\Delta\Omega^{(0)}_{\text{site}} 
                           + \Delta\Omega^{(1)}_{\text{site}} \, .
\end{equation}

Using definition (\ref{eq:Omega}) of the grand-canonical potential, we
calculate the above changes from the restricted partition functions of
rooted trees:
\begin{align}
  e^{-\mu\Delta\Omega_{\text{link}}} & =
  \frac{\overbrace{\Xi_e^{(1)}(\Xi_e^{(1)})^p+\Xi_*^{(1)}(\Xi_e^{(0)})^p+
    p\,\Xi_e^{(1)}\Xi_*^{(0)}(\Xi_e^{(0)})^{p-1}}^
{\text{connected rooted trees}}}
{\underbrace{(\Xi_e^{(1)}+\Xi_*^{(1)})(\Xi_e^{(0)}+\Xi_*^{(0)})^p}
  _{\text{disconnected rooted trees}}} \nonumber \\
    & = \frac{1+x^{(1)}+p\,x^{(0)}}{(1+x^{(1)})(1+x^{(0)})^p}\, ,
\end{align}
where we have divided both numerator and denominator by
$\Xi_e^{(1)}(\Xi_e^{(0)})^p$. Further on we get
\begin{align}
  e^{-\mu\Delta\Omega^{(1)}_{\text{site}}} & =
    \frac{\overbrace{e^{\mu}(\Xi_e^{(0)})^{p(k+1)}}^{\text{1-vertex 
         occupied}}+\overbrace{\bigl[(\Xi_e^{(0)})^p+p\,\Xi_*^{(0)}
     (\Xi_e^{(0)})^{p-1}\bigr]^{k+1}}^{\text{1-vertex empty}}}
         {\underbrace{(\Xi_e^{(0)}+\Xi_*^{(0)})^{p(k+1)}}_{\text{free 
               rooted trees}}}  
\nonumber \\
    & = \frac{e^{\mu}+(1+p\,x^{(0)})^{k+1}}{(1+x^{(0)})^{p(k+1)}}\, ,
\end{align}
and
\begin{align}
   e^{-\mu\Delta\Omega^{(0)}_{\text{site}}}
     & = \frac{e^{\mu}\bigl[\Xi_e^{(1)}(\Xi_e^{(0)})^{p-1}\bigr]^{k+1}+
           \bigl[\Xi_e^{(1)}(\Xi_e^{(0)})^{p-1}+ 
                  \Xi_*^{(1)}(\Xi_e^{(0)})^{p-1}+
                  (p-1)\Xi_e^{(1)}\Xi_*^{(0)}
                 (\Xi_e^{(0)})^{p-2}\bigr]^{k+1}}
              {\bigl[(\Xi_e^{(1)}+\Xi_*^{(1)})
               (\Xi_e^{(0)}+\Xi_*^{(0)})^{p-1}\bigr]^{k+1}} \nonumber
             \\
   & = \frac{e^{\mu}+\bigl(1+x^{(1)}+(p-1)x^{(0)}\bigr)^{k+1}}
            {\bigl[(1+x^{(1)})(1+x^{(0)})^{p-1}\bigr]^{k+1}} \, .
\end{align}
This allows us to calculate the grand-canonical potential for the
different crystalline and liquid solutions for the cavity fields, and
thus to determine the phase diagram of the model by looking for the
locally stable solution of minimal potential -- provided the
partitioning of the lattice in the two sublattices, i.e. the
assumption of a crystalline structure, is consistent with the global
structure of the lattice.

\subsection{The phase diagram}
\label{sec_phasendia}

Let us shortly summarize the case $p=1$ where we have an ordinary
Bethe lattice. We have found that at $\rho_{cr}=\frac{1}{k+1}$ a
crystalline solution appears continuously from the liquid one, i.e. we
find a second-order crystallization transition. For $\mu<\mu_{cr}$,
with $\rho(\mu_{cr}) = \rho_{cr}$, the liquid solution is the only
one, and thus globally stable. For $\mu>\mu_{cr}$, this solution
becomes locally unstable, and two stable crystalline states
exist. They are connected to each other by exchanging the two
sub-lattices. For $\mu\to\infty$, the system approaches one of its
ground states which is given by a periodic alternation of occupied and
empty sites.

In the case of a generalized Bethe-lattice, i.e. for $p>1$, the
situation is slightly more involved. Below the spinodal point
$\mu_{sp}$ only the liquid solution exists, it thus describes the
globally stable solution. It exists, however, also for larger chemical
potentials and, up to $\mu_{cr}$ (as defined in the last paragraph),
it remains at least locally stable under small perturbations.

At $\mu_{sp}$ two other, crystalline solutions appear. The first one
has always $\rho^{(1)}>\rho^{(0)}$, and the global density is larger
than the liquid one. This solution is always locally stable, as can be
shown by analyzing the relaxational dynamics of the system
\cite{self}. The second crystalline solution starts with the same
sub-lattice densities at the spinodal point, but its global density
first decreases. In addition it is initially locally unstable, hence
unphysical. At $\mu=\mu_{cr}$, however, both sub-lattice densities
become equal, $\rho^{(1)}=\rho^{(0)}=\frac{1}{k+1}$, and exactly at
this point it thus coincides with the liquid solution. Beyond
$\mu_{cr}$, the new solution becomes locally stable, and thus takes
over the local stability from the liquid solution.  Since both
$\rho(\mu)$-curves touch in this point with the same slope but
different curvature, the corresponding transition there is of {\it
third order}. The crystalline solution becomes also in some other
aspect quite particular: The sub-lattice densities now fulfill
$\rho^{(1)}<\rho^{(0)}$, i.e. every clique has $p$ vertices of higher,
and one vertex of lower density! Note that this transition exists only
if the global density $\frac{1}{k+1}$ can be reached. Every clique can
be occupied by at most one particle, the density is thus restricted to
values $\rho<\frac{1}{p+1}$. A necessary and sufficient condition for
the existence of the inverse crystalline solution is thus given by
$k>p$.

For all $\mu>\mu_{sp}$ we thus find two locally stable
solutions, and we have to consider the grand-canonical potentials in
order to decide which solution is the globally stable one. For $p>1$,
the liquid solution remains the thermodynamically relevant one up to
some $\mu_s>\mu_{sp}$, above which the first crystalline solution
has the minimal $\omega$; a first-order freezing transition occurs.
The second crystalline solution has never the globally minimal
grand-canonical potential, but it crosses the one of the liquid
solution at $\mu_{cr}$, verifying thus the results of the local
stability analysis.

It has to be remarked that in the case of ordinary Bethe-lattices,
i.e. $p=1$, the equations are symmetric with respect to $\rho^{(1)}$
and $\rho^{(0)}$, thus normal and inverse crystallization are
indistinguishable. We can therefore consider this case as a degenerate
one, where the crystalline and the inverse crystalline solutions
appear in the same moment, and thus $\mu_{sp}=\mu_{s}=\mu_{cr}$ holds.

\subsection{An example}

We want to illustrate this behavior in a specific example. The system
of minimal $k$ and $p$ showing inverse crystallization has $k=3$ and
$p=2$. We therefore concentrate on these values.

\begin{figure}[htbp]
  \begin{center}
    \includegraphics[height = 9cm]{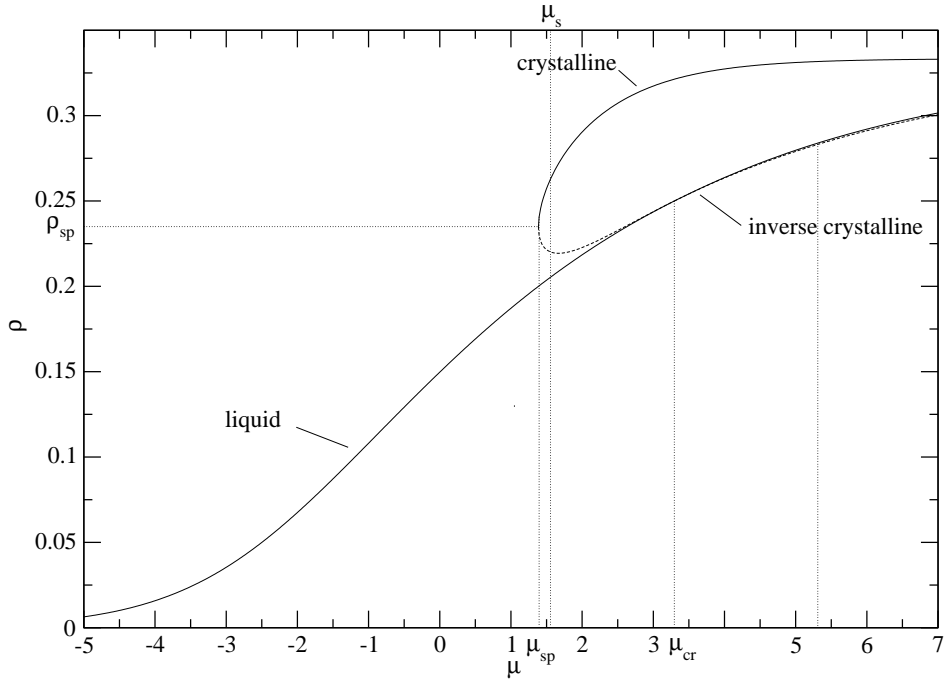}
    \caption{Phase diagram for a generalized Bethe-lattice with $k=3$,
      $p=2$. At $\mu_{sp}$ a crystalline solution sets in
      discontinuously, and at $\mu_s$ it becomes globally stable.  At
      $\mu_{cr}$ the liquid solution becomes locally
      unstable, and goes over into an inverse crystalline phase, which
      is relevant for $\mu_{cr} < \mu < \mu_{cg}$ with $\mu_{cg}\le 5.31$. Locally stable solutions are denoted by full lines,
      locally unstable ones by dashed lines.}
\label{fig_rhostat}
  \end{center}
\end{figure}

\begin{figure}[htbp]
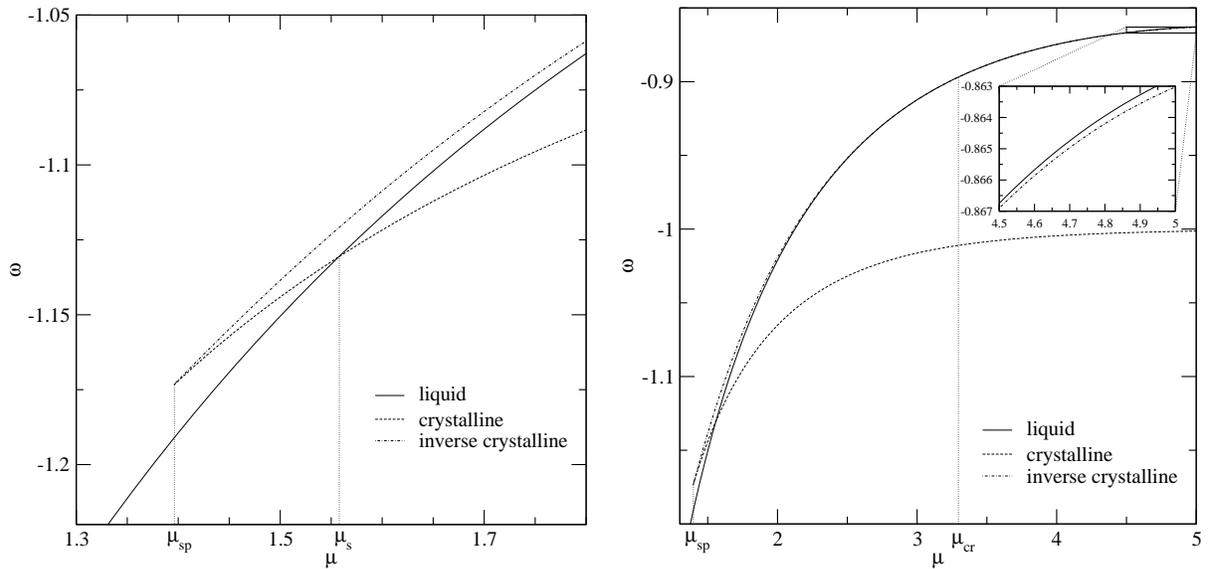

  \begin{center}
    \begin{center}
    \includegraphics[height = 7.5cm]{grpotlicr1.eps}
    \quad
    \includegraphics[height = 7.5cm]{grpotlicr2.eps}
    \end{center}
    \caption{Grand-canonical potential per clique $\omega$ for $k=3$,
    $p=2$.}
\label{fig_grosspot}
  \end{center}
\end{figure}

The phase diagram is shown in Fig.~\ref{fig_rhostat}, the
grand-canonical potential per clique is shown in
Fig.~\ref{fig_grosspot}. For small $\mu$, only the liquid phase
exists. At $\mu_{sp}\simeq 1.386$ a crystalline solution with
sub-lattice densities $\rho^{(0)}_{sp}\simeq 0.034$,
$\rho^{(1)}_{sp}\simeq 0.637$ and global density $\rho_{sp}\simeq
0.235$ appears. Up to $\mu_s\simeq 1.558$, the liquid phase remains
however the thermal equilibrium of our model. For all $\mu$ above
$\mu_s$, the crystalline solution of higher density becomes the
globally stable one, i.e. at $\mu_s$ the systems undergoes a
first-order crystallization transition. The liquid solution stays,
however, also locally stable beyond $\mu_s$, and in absence of
sufficiently strong perturbations the system remains liquid also for
higher $\mu$. Only at
$\mu_{cr}\simeq 3.296$ it loses its local stability in a third-order
transition to an inverse crystalline state with
$\rho^{(0)}>\rho^{(1)}$.  Note that the density curves of the liquid
and the second crystalline solution only touch at $\mu_{cr}$, the
inverse crystalline density is slightly larger than the liquid one,
whereas the grand-canonical potentials cross each other with equal
slopes.

At this point we have to remark that the entropy $s=-\mu(\omega+\rho)$
of the inverse crystalline solution vanishes at $\mu \simeq 5.58$, and
it becomes therefore unphysical at higher chemical potentials. Using a
slight generalization of the 1RSB formalism introduced in the next
section, we find for $\mu > 5.31$ the existence of a solution that
corresponds to a {\it crystalline glass phase} (cf. section
\ref{sec_cryglass}) which seems appropriate to solve the entropy
crisis. Thus, the inverse crystallization is relevant only in an
interval $\mu_{cr}<\mu<\mu_{cg}$ with $\mu_{cg} \le 5.31$.    

\subsection{Comparison to numerical experiments}
\label{sec:compnumerics}
To test our analytical results on the crystallization transitions, we
have performed Monte-Carlo simulations. These are naturally performed
on finite lattices, and a natural finite analog of the Bethe lattice
defined as an infinite tree is provided by regular random graphs. As
already discussed in Sec.~\ref{sec:model}, their large scale structure
contains random loops which are not compatible with any division into
0- and 1-sub-lattices. This can be circumvented by imposing the
consistency with the crystalline structure by generating, e.g., a
random regular bipartite graph by grouping the vertices into two
subsets of sizes $N$ and $pN$, and by introducing cliques only
in between one vertex from the first and $p$ vertices from the second
subset.

Note that also these graphs have an important difference to the
generalized Bethe-lattices defined as infinite graphs which survives
even in the thermodynamic limit: The first graphs allow only for
exactly one partition into 0- and 1-lattices, whereas the second
definition allows for an exponential number of such
partitions. This difference is meaningful for stability properties
of the system. It can be shown in numerical simulations that even when
going to $(p+1)$-partite graphs, that allow for $p+1$ partitions,
the inverse crystalline phase becomes unstable.


In the MC simulations we have included the following microscopic
processes: Particles are mobile, i.e.~they are allowed to move to
neighboring sites whenever these are empty and do not have further
occupied neighbors (hard-core constraint), and the system is coupled
to a particle bath such that particles can be generated or deleted.
The rates for the last two processes have to be related by detailed
balance, i.e. they have ratio $e^\mu$, in order to guarantee
equilibration.

\begin{figure}[htbp]
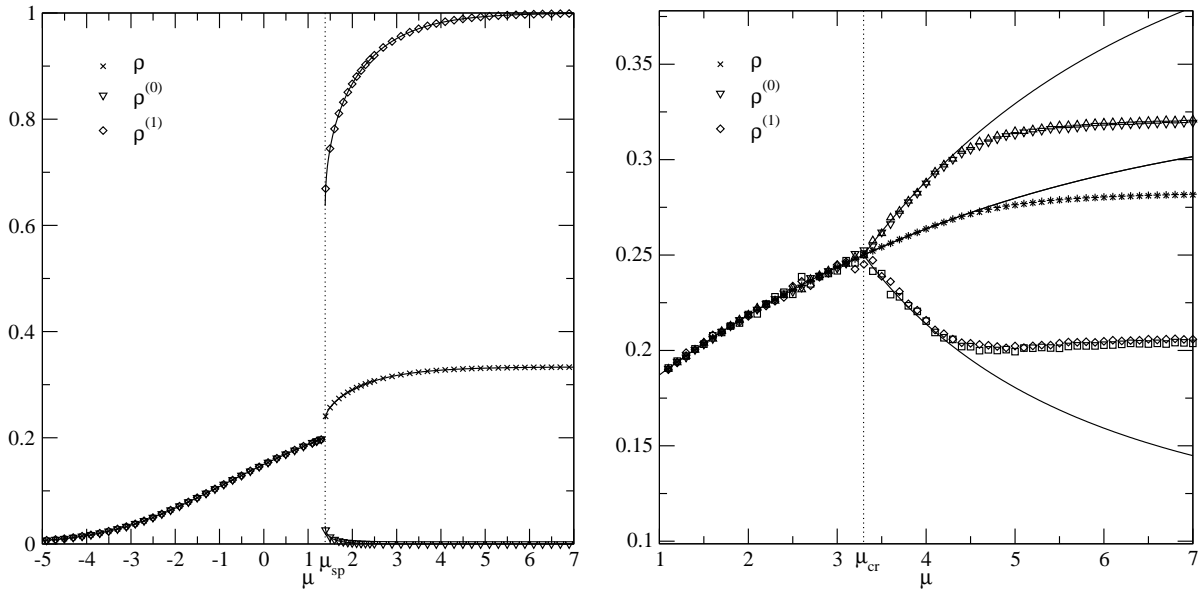

  \begin{center}
    \begin{center}
    \includegraphics[height = 7.8cm]{decomp.eps}
    \quad
    \includegraphics[height = 7.75cm]{inverseUp.eps}
    \end{center}
    \caption{Decompactification experiment for normal crystallization
    (left figure) and compactification experiment for inverse
    crystallization (right figure). Symbols 
    which are rotated with respect to the legend correspond to smaller
    rates $|\delta \mu / \delta t|$, see main text. Analytical results
    are represented by full lines.}
\label{fig_comp}
  \end{center}
\end{figure}

First, we have tested our predictions for the normal crystal phase on
large graphs of $N=6\cdot10^6$ vertices. We have started the
experiment in one densest packing, and then decreased the chemical
potential with a rate of $\delta \mu / \delta t = -0.2/(100\text{
MCs})$, and close to $\mu_{sp}$ with $\delta \mu / \delta t = -
0.1/(200\text{ MCs})$, starting with $\mu=7$. A further decrease in
the decompactification rate did not alter the results, so we can
expect thermal equilibrium. The left part of Fig.~\ref{fig_comp}
shows, that the results are in perfect agreement with the analytical
ones, in particular the discontinuous decrease of the density at the
spinodal point is demonstrated with very high precision.

The results for the inverse crystallization were tested by starting
with an empty system which was compactified in a first step to $\mu=1$
where only the liquid phase and no crystalline phases exist. The
system was compactified by increasing $\mu$ in discrete steps with two
very small rates, $\delta \mu / \delta t =
0,1/(10\,000\text{ MCs})$ and $\delta \mu / \delta t =
0,1/(5\,000\text{ MCs})$. Due to this slow dynamics we have also used
a slightly smaller system of $N=3\cdot10^6$ vertices.

The results are represented in the right of Figs.~\ref{fig_comp}. The
first observation is that, for small $\mu$, numerical and analytical
data coincide impressively: We find the liquid phase with equal sub-lattice
densities for $\mu<3.3$ and a transition to the inverse crystalline
phase at $\mu_{cr}\simeq 3.3$, as predicted in the last section. 
For $\mu> \mu_{cg} \simeq 4.1$, the results depend on the
compaction rate, underlining the glassy character of the corresponding
phase. It is also observed that the global density is smaller than the
inverse crystalline one, which mainly results from a lower sublattice
density on the 0-lattice.

We mention that besides the transition from the liquid to the inverse
crystalline phase which is shown in Fig.~\ref{fig_comp} (right), there
is a certain probability for the system to undergo a transition to the
normal crystalline phase with $\rho^{(1)}>\rho^{(0)}$ when compacted
beyond $\mu_{cr}$. The dependence of this probability on the
compaction rate and on the system size might be an interesting
subject of further investigation.  

\section{Glassy behavior}
\label{chap_1RSB}

In the last section, we have already seen that the simple crystalline
ansatz of two cavity fields $x^{(0)}$ and $x^{(1)}$ for solving the
iterative Eq.~\ref{gl_lokfeldrek} leads to inconsistencies for large
$\mu$. The entropy of the inverse crystalline solution, which
describes a metastable phase, becomes negative at finite $\mu$, and
the solution is not physical any more. Frequently zero-entropy points
can be understood as an indicator for the existence of a glass
transition at some smaller value of the chemical potential, and thus
a more general solution ansatz has to be considered.

The situation is even more drastic if we consider random regular
graphs as realizations of Bethe lattices. As already discussed, they
are locally isomorphic to a tree, but they contain random large-scale
loops which are not consistent with any crystalline structure. In
particular they do not allow for a coherent partition of the original
graph into two subgraphs as discussed in the context of the
crystalline solution, the system is frustrated. So far, for these
graphs we have only the liquid solution which, however, is in
contradiction to numerical results for large $\mu$.

We therefore have to consider the possibility of the existence of a
glassy high-density phase, or, in technical terms, we have to search
for a replica symmetry broken solution (RSB). On the level of one-step
RSB (1RSB) this can be achieved most easily using the cavity approach
\cite{mezparvir}, as developed for finite-connectivity systems in
\cite{mezpar}, and used also in \cite{rivoire} for a similar lattice
gas model.

We first discuss a homogeneous 1RSB solution, which is valid in
particular for lattices not allowing for crystallization. At the end
of this section, we also show shortly which modifications have to be
made in order to account for the crystalline glass phase which was
already mentioned at the end of the last section.

\subsection{The 1RSB cavity solution}
\label{sec_cavity}

We want to construct a non-periodic solution of
Eq.~(\ref{gl_lokfeldrek}). For small $\mu$, the lattice gas is very
diluted, so the liquid solution is expected to be correct. For higher
densities, the frustration in the systems becomes, however, relevant.
In this regime, Eq.~(\ref{gl_lokfeldrek}) is expected to have an
exponential number of solutions, each one describing a thermodynamic
{\it state} of the model. These states are, in particular, not
spatially homogeneous. Let us therefore assume that the number
$\mathcal{N}_N(\omega)$ of states of density $\omega$ of the
grand-canonical potential satisfies
\begin{equation}
\label{eq:complexity}
  \mathcal{N}_N(\omega) \sim e^{N\Sigma(\omega)} \, ,
\end{equation} 
where $\Sigma(\omega) \ge 0$ is the so-called {\it complexity} (or
{\it configurational entropy}). It is assumed to be a growing concave
function.

We consider now a rooted tree with root $jl$ as shown in
Fig.~\ref{fig_betherek} for $j=0$, $l=0$, and assume that the
solutions for the cavity field $x^{jl}$ and the potential
$\omega^{jl}$ are distributed according to some probability
distribution $R^{jl}(x^{jl}, \omega^{jl})$. Repeating the iteration
step represented in Fig.~\ref{fig_betherek}, we can calculate the
distribution for the root from the distribution of the neighboring roots,
\begin{align} \label{gl_distrekinhom}
  R^{00}(x^{00}, \omega^{00}) & = \int \prod_{j=1}^{k}\prod_{l=1}^{p}
  dx^{jl} d\omega^{jl} R^{jl}(x^{jl}, \omega^{jl})
  \,\delta\bigl(x^{00}-\hat{x}(x^{11}, \ldots , x^{kp})\bigr) \nonumber \\
  & \qquad \times \delta\Biggl(\omega^{00}- 
 \frac{1}{N^{00}}\biggl(\sum_{j=1}^{k}\sum_{l=1}^{p}N^{jl}\omega^{jl}
    +\Delta\Omega_{\text{iter}}(x^{11}, \ldots , x^{kp}) \biggr)
    \Biggr) \, .
\end{align}
We have used the number $N^{jl}$ of the vertices in the corresponding
rooted tree, i.e.
\begin{equation}\label{gl_wbnumbersum}
  N^{00} = \sum_{j=1}^{k}\sum_{l=1}^{p}N^{jl} + 1 \, .
\end{equation}
The change in the grand canonical potential is calculated via
\begin{eqnarray}\label{gl_omiter}
  e^{-\mu\Delta\Omega_{\text{iter}}} & = &
  \frac{\text{partition function after iteration}}{
    \text{partition function before iteration}}
 \nonumber \\ 
  & = &
  \frac{\Xi_e^{00}+\Xi_*^{00}}
    {\prod_{j=1}^{k}\prod_{l=1}^{p}
        \bigl(\Xi_e^{jl}+\Xi_*^{jl}\bigr)} \nonumber \\
  & \overset{(\ref{gl_rekzsumme})}{=} &
      \frac{\prod_{j=1}^{k}\bigl(\prod_{l=1}^{p}\Xi_e^{jl}+
        \sum_{l=1}^{p}\Xi_*^{jl}\prod_{m\ne l}\Xi_e^{jm}\bigr)+
         e^{\mu}\prod_{j=1}^{k}\prod_{l=1}^{p}\Xi_e^{jl}}
       {\prod_{j=1}^{k}\prod_{l=1}^{p}
        (\Xi_e^{jl}+\Xi_*^{jl})} \nonumber \\
  & = & \frac{\prod_{j=1}^{k}\bigl(1+\sum_{l=1}^{p}x^{jl}\bigr)+e^{\mu}}
      {\prod_{j=1}^{k}\prod_{l=1}^{p}\bigl(1+x^{jl}\bigr)} \, .
\end{eqnarray}

Note that Eq.~(\ref{gl_distrekinhom}) assumes factorization of the {\it
joint} distribution $ \tilde{R}(x^{11}, \omega^{11}; \ldots ; x^{kp},
\omega^{kp})$ of all neighbors of vertex $00$. This would obviously be
correct on a tree, where different rooted subtrees are disconnected
and thus statistically independent. On a (generalized) random regular
graph, these vertices are not the roots of $kp$ disconnected rooted
trees, so the factorization is not {\em a priori} true. Loops have, however,
a typical length of ${\cal O}(\ln N)$, so once their short connection
via vertex $00$ is disregarded, they are far away. On the level of
1RSB we therefore assume that the joint distribution factorizes, and
we can use Eq.~(\ref{gl_distrekinhom}).

On these graphs, regular structures like a crystalline one are
forbidden by the existence of random loops. We therefore assume
homogeneity on the level of the field distributions, $R^{00} \equiv
R^{11} \equiv \ldots \equiv R^{kp} \equiv R$, thus
Eq.~\ref{gl_distrekinhom} becomes a self-consistent equation for
$R(x,\omega)$. In order to solve this equation, we use assumption
(\ref{eq:complexity}) and write
\begin{equation}\label{gl_fakRansatz} 
  R(x, \omega) \sim e^{N\Sigma(\omega)}\,P^{(\omega)}(x)\, , \quad
  \text{with} \quad
  \int dx P^{(\omega)}(x) = 1 \, ,
\end{equation}
where $P^{(\omega)}$ is the distribution of cavity fields with fixed
$\omega$. In order to derive an equation for this function, we fix the
density of the grand-canonical potential at some arbitrary value
$\omega_0$, and introduce the parameter
\begin{equation}\label{gl_mom0def}
  m(\omega_0) \doteq
  \frac{1}{\mu}\,\frac{d\Sigma}{d\omega}(\omega_0)
\end{equation}
and linearize the complexity in a small vicinity
$\mathcal{V}_{\omega_0}$ of $\omega_0$:
\begin{equation}\label{gl_sigmalin}
  \Sigma(\omega) \simeq \Sigma(\omega_0) + m\mu(\omega-\omega_0) \, .
\end{equation}

Since $\Sigma$ was assumed to be concave, there is a one-to-one
correspondence between $\omega_0$ and $m$. We define the distribution
\begin{equation}\label{gl_defPm}
  P^{(m)}(x) = \frac 1{|\mathcal{V}_{\omega_0}|} 
  \int_{\omega\in\mathcal{V}_{\omega_0}} d\omega P^{(\omega)}(x) \, .
\end{equation}
Plugging this into Eq.~(\ref{gl_distrekinhom}), we find in the limit
of a vanishing vicinity $\mathcal{V}_{\omega_0}$ the self-consistent
equation
\begin{equation}\label{gl_1RSBcavity}
  P^{(m)}(x) \propto \int \prod_{j=1}^{k}\prod_{l=1}^{p}
  dx^{jl} P^{(m)}(x^{jl})\,\delta(x-\hat{x}(x^{11}, \ldots , x^{kp}))
  \,e^{-m\mu\Delta\Omega_{\text{iter}}(x^{11}, \ldots , x^{kp})}
\end{equation}
which is called the {\it factorized 1RSB cavity equation} for the
cavity field distribution in our hard-sphere lattice gas model.

To clarify the meaning of $m$, we calculate
\begin{equation}\label{gl_Ximdef}
  \Xi(m)  = \sum_{\alpha} e^{-Nm\mu\omega^{(\alpha)}}
   = \int d\omega\,\mathcal{N}_N(\omega) e^{-Nm\mu\omega}
   \sim \int d\omega\,e^{N(\Sigma(\omega)-m\mu\omega)}
   = e^{-Nm\mu\Phi(m)}
\end{equation} 
with the sum running over all states $\alpha$, i.e. over all solutions
of Eq.~(\ref{gl_lokfeldrek}). Note that, for $m=1$, this reproduces
the grand-canonical partition function. In the thermodynamic limit,
$\Phi(m)$ is given by the saddle-point approximation,
\begin{equation} \label{gl_sattpktPhim}
  \Phi(m) = \omega - \frac{1}{m\mu}\Sigma(\omega) \, , \qquad
  \text{with} \qquad
  \partial_{\omega}\Sigma(\omega)=m\mu \, ,
\end{equation}
which reproduces the definition of $m(\omega)$ in Eq.~(\ref{gl_mom0def}).
From Eq.~(\ref{gl_sattpktPhim}) follows that
\begin{equation}\label{gl_partialmphi}
  \partial_m\Phi(m) = \partial_m\omega +
  \frac{1}{m^2\mu}\Sigma(\omega) -
  \frac{1}{m\mu}\partial_{\omega}\Sigma(\omega)\partial_m\omega
   =  \frac{1}{m^2\mu}\Sigma(\omega) \, .
\end{equation}

Let us compare Eq.~(\ref{gl_sattpktPhim}) with the usual equation for
the entropy
\begin{equation}
    \Phi(m) = \omega - \frac{1}{m\mu}\Sigma(\omega) \qquad \qquad
    \omega = -\rho - \frac{1}{\mu}\,s \, .
\end{equation}
There is an obvious similarity: As $s$ counts the number of
microscopic configurations, $\Sigma$ counts the number of states. As
$\mu$ serves as a control parameter selecting configurations of a
given density $\rho$, the parameter $m$ allows to focus on states of
arbitrarily given $\omega$. Therefore $\Phi(m)$ plays, on the level of
states, the same role as the grand-canonical potential on the level of
configurations. Following this analogy, we can calculate $\Phi(m)$ as
done in Sec.~\ref{sec_bethepeierls} for the grand-canonical potential:
\begin{equation}\label{gl_bethepeiphi}
      \Phi(m) = -\frac{p\,(k+1)}{p+1}\, \Delta\Phi_{\text{link}}(m) +
  \Delta\Phi_{\text{site}}(m) 
\end{equation}
with
\begin{equation}\label{gl_1RSBdpsite}
  e^{-m\mu\Delta\Phi_{\text{site}}} = \int
        \prod_{j=1}^{k+1}\prod_{l=1}^{p} dx^{jl} P(x^{jl})
        e^{-m\mu\Delta\Omega_{\text{site}}(x^{11}, \ldots , x^{k+1,
        p})}
\end{equation}
and
\begin{equation}\label{gl_1RSBdplink}
 e^{-m\mu\Delta\Phi_{\text{link}}} = \int \prod_{l=1}^{p+1}dx^{(l)}
        P(x^{(l)}) e^{-m\mu\Delta\Omega_{\text{link}}(x^{(1)}, \ldots
        , x^{(p+1)})}\ .
\end{equation}
$\Delta\Omega_{\text{site}}$ and $\Delta\Omega_{\text{link}}$ are the
direct generalizations of the corresponding quantities in
Sec.~\ref{sec_bethepeierls}:
\begin{equation}\label{gl_appdelomallg}
  e^{-\mu \Delta\Omega_{\text{link}}} = 
    \frac{1+\sum_{l=1}^{p+1} x^{(l)}}{\prod_{l=1}^{p+1}(1+x^{(l)})} \,
    , \qquad
  e^{-\mu \Delta\Omega_{\text{site}}} = 
    \frac{e^{\mu}+\prod_{j=1}^{k+1}(1+\sum_{l=1}^{p}
      x^{jl})}{\prod_{j=1}^{k+1}\prod_{l=1}^{p}(1+x^{jl})} \, .
\end{equation}

\subsubsection{Static and dynamic glass transition}

Let us consider the 1RSB equation (\ref{gl_1RSBcavity}) first for $m=1$.
As already mentioned, $\Xi(m=1)$ then gives the usual partition function,
\begin{equation}\label{gl_intzusum}
  \Xi(\mu) \sim \int d\omega e^{N(\Sigma(\omega)-\mu\omega)} \, ,
\end{equation}  
where the integration runs over the interval $(\omega_{\text{min}},
\omega_{\text{max}})$ with positive $\Sigma(\omega)$.

Let us first discuss the case $p>1$.  For small $\mu$,
Eq. (\ref{gl_1RSBcavity}) has only the liquid solution $P(x) =
\delta(x-x^*)$. The particle density is low enough that the
frustration resulting from long loops is not relevant.  At some
chemical potential $\mu=\mu_d$, the first non-trivial solution appears
discontinuously. This point is called the {\it dynamical glass
transition}.  For $\mu>\mu_d$ the system has exponentially many states,
the complexity $\Sigma(\omega_s)>0$ is positive. The grand-canonical
potential $\omega_s$ of one state can be calculated via the saddle
point method from Eq. (\ref{gl_intzusum}), i.e. as a solution of $\mu =
\partial_{\omega}\Sigma(\omega)$. The potential $\omega^*$ of the
liquid solution is given by $\omega^*=\Phi(m=1)$. With
Eq. (\ref{gl_sattpktPhim}) we conclude $\omega_s>\omega^*$. The states
born at $\mu_d$ are therefore metastable ones, the thermodynamically
stable solution remains still the liquid one. The dynamics of the
system becomes, however, trapped already at $\mu_d$ by the exponential
number of metastable states, i.e. the system remains dynamically out
of equilibrium, cf. the discussion in \cite{monasson}.

Increasing $\mu$ beyond $\mu_d$, the complexity $\Sigma(\omega_s)$
decreases, until it vanishes at $\mu=\mu_s$.  According to
Eq. (\ref{gl_sattpktPhim}), the grand-canonical potential $\omega_s$
of the 1RSB solution equals here the one of the liquid solution, and
$\mu=\mu_s$ indicates thus the location of a {\it static},
i.e. thermodynamic glass transition.

For $\mu>\mu_s$ the naive application of the saddle-point method leads
to an unphysical solution $\omega_s<\omega_{\text{min}}$ of negative
complexity. The correct exponentially dominating contribution in
Eq. (\ref{gl_intzusum}) is thus given at the interval limit $\omega_s
= \omega_{\text{min}}$. In this situation we need a ``corrected''
saddle-point equation $m\mu=\partial_{\omega}\Sigma(\omega)$
(cf.~Eq.~(\ref{gl_sattpktPhim})), which defines the value $m_s =
\frac{1}{\mu}\partial_{\omega}\Sigma(\omega=\omega_s)$. According to
Eq.~(\ref{gl_sattpktPhim}) we thus obtain the grand-canonical
potential of the 1RSB equilibrium via $\Phi(m_s) = \omega_s =
\omega_{\text{min}}$. Using $\Sigma(\omega_s) =
\Sigma(\omega_{\text{min}}) = 0$ and Eq.(\ref{gl_partialmphi}) we thus
find
\begin{equation}\label{gl_wahlms}
  \partial_m\Phi(m=m_s) = 0 \, .
\end{equation}

Also other values of $m$ have a physical significance. For $m<m_s$ we
get metastable states with $\omega>\omega_s$. These are important for
understanding the dynamics of the system. A particular importance has
here the value $\omega_d$ where the complexity is maximized,
cf. Sec~\ref{sec_closepack}. Larger values $m>m_s$ lead to negative
complexity which can be reinterpreted in the context of atypical
graphs - their complexity corresponds to the probability that a
(generalized) random regular graph has states of this smaller
grand-canonical potential. An example for such an untypical graph is
the multi-partite one allowing for a crystalline solution as discussed
before.

The situation is easier for $p=1$, i.e. for standard Bethe
lattices. Here $\mu_d$ and $\mu_s$ coincide and mark a continuous
spin-glass transition. The region, where the dynamics is dominated by
an exponential number of metastable states, disappears.

\subsubsection{The spin-glass instability of the liquid solution}  

An upper bound for $\mu_d$ is given by the spin-glass instability of
the liquid solution. This instability means that a small perturbation
of $x^{*}$ is amplified during the iteration of
Eq.~(\ref{gl_1RSBcavity}). The criterion for the instability is
\begin{equation} \label{gl_Spinglasinst}
  k p \left|\frac{\partial\hat{x}(x^{*}, \ldots, x^{*})}{\partial
      x^{11}}\right|^2>1\, . 
\end{equation}

Equality holds at critical values $\mu_g$ of the chemical potential,
and $\rho_g$ of the density. According to Eq.~(\ref{gl_lokfeldrek})
this happens at
\begin{equation}\label{gl_rhog}
 \rho_g = \frac{1}{\sqrt{k p\,}+1} \, .
\end{equation} 
Above this point, the liquid solution is locally unstable.

For $p=1$, also this instability coincides with $\mu_d$ and $\mu_s$.

\subsubsection{Solving the 1RSB equation via population dynamics}

Solving Eq.~(\ref{gl_1RSBcavity}) seems to be too complicated, we
therefore have to solve it numerically. The main idea hereby is to
iterate the self-consistency equation until a fixed point is reached.
This happens by representing first $P^{(m)}$ by a finite but large
population $\{x_1, \ldots , x_M\},\, M\gg1$ of cavity fields, and by
updating this population via a population dynamical algorithm
\cite{mezpar}: In every step $kp$ fields are selected randomly from
the population, and a new field is calculated by $\hat x$. This field
replaces one or more of the old fields in the population, where the
number of substituted fields is a function of the reweighting factor
in Eq.~(\ref{gl_1RSBcavity}).

For our example $k=3,\, p=2$, we find the following results: From
Eq.~(\ref{gl_rhog}) we conclude that $\rho_g = \frac{1}{\sqrt{6}+1}
\simeq 0.290$. It is, in particular, larger than
$\rho_{\text{krit}}=\frac{1}{k+1}=0.25$, which is true for all $k>p$.
With Eq.~\ref{gl_lsgstatikfl} we find $\mu_g\simeq 5.89$, this value
is confirmed by the population-dynamical algorithm.

Further on, we find that at $\mu_d \simeq 4.85$ the first non-trivial
solution for $P^{(m)}$ emerges discontinuously. In principle also the
static spin-glass transition $\mu_s$ can be determined by the
population-dynamical algorithm. The problem is, however, that it
results from a comparison of the grand-canonical potentials of the
liquid and the glassy solutions, which are subject to large
fluctuations. Even a large statistics allows only for a rough estimate
of $\mu_s\simeq 5.0$.

\begin{figure}[htbp]
  \begin{center}
    \begin{center}
    \includegraphics[height = 7.8cm]{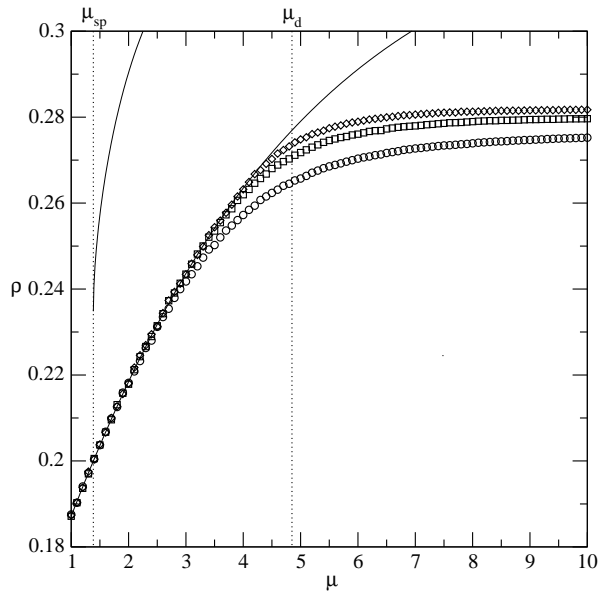}
    \end{center}
    \caption{Compactification experiments for a generalized random
    regular graph $k=3,\, p=2$, and $N=450 000$ sites. The full lines
    give the analytical results for the liquid and crystalline
    densities, whereas the symbols correspond to compaction rates
    $\delta \mu/\delta t = 0.1/10$ MCs (circles), $0.1/100$ MCs
    (squares) and $0.1/1000$ MCs (diamonds). Obviously, the systems
    falls out of the liquid equilibrium when the equilibration time
    starts to exceed the waiting time at the corresponding chemical
    potential, i.e. before the latter reaches $\mu_d$.}
\label{fig:glasstrans}
  \end{center}
\end{figure}

Note that, for the case considered here, the dynamical and the static
RSB transition are located beyond the inverse-crystallization point,
where the liquid solution becomes locally unstable. It is, however,
possible to validate the correctness of the result by numerical
simulations which have to be performed on a (generalized) random
regular graph. These graphs have a large-scale loop structure which is
not consistent with any crystalline packing, cf.~the discussion in
Sec.~\ref{sec:model}, and thus the system undergoes directly the glass
transition if compactified until $\mu_d$. As can be seen in
Fig.~\ref{fig:glasstrans}, the compaction dynamics falls out of
equilibrium even before the dynamical glass transition, but slower and
slower compaction allows for a closer approach to $\mu_d$, showing
thus the compaction-rate effects which are typical for glass formers.

\subsubsection{Glass-instability of the inverse crystalline phase
  ($k=3$, $p=2$)}
\label{sec_cryglass}
A stability analysis of the inverse crystalline phase toward a 
1RSB ``glassy'' phase requires to embed the inverse crystalline
solution in the 1RSB formalism. To this purpose we rewrite the 1RSB
equation \eqref{gl_1RSBcavity} without assuming homogeneity of the field
distributions $P^{00}, P^{11}, \ldots , P^{kp}$:
\begin{equation}
  P^{00}(x) \propto \mathcal{F}[P^{11}, \ldots , P^{kp}](x)
\end{equation}
where $\mathcal{F}$ contains the r.h.s. of the 1RSB equation. The
dependence on $m$ is eliminated by putting $m \doteq 1$ which can be
justified as we are only interested in transitions points to the
glassy phase. The canonical embedding now reads
\begin{equation}\label{gl_1RSBinvcr}
  P^{(0)} \propto \mathcal{F}[P^{(1)}, P^{(0)}, \ldots , P^{(0)}; \ldots]\,
  , \quad P^{(1)} \propto \mathcal{F}[P^{(0)},  \ldots , P^{(0)}; \ldots] \, ,
\end{equation}
where the displayed blocks in the argument of $\mathcal{F}$ must be
repeated $k$ times similar to Eq. \eqref{gl_lfrekallg}. The distributions
$P^{(0)}$ and $P^{(1)}$ are simply $\delta$-functions at the values
$x^{(0)}$ and $x^{(1)}$ of the local fields according to the inverse
crystalline solution.
The local instability of the iteration defined by
Eq. \eqref{gl_1RSBinvcr} can be
detected by evaluating $\mathcal{F}$ for $P^{(0)}$ and $P^{(1)}$
chosen as Gaussian distributions with respective variances $\varepsilon^{(0)}$
and $\varepsilon^{(1)}$. Under neglecting of the reweighting factors
in $\mathcal{F}$ one easily calculates the variances
${\varepsilon^{(0)}}'$ and ${\varepsilon^{(1)}}'$ after one step of the
iteration:
\begin{eqnarray} \label{gl_eps01it}
   {\varepsilon^{(0)}}' & = & 
   \sqrt{k(p-1)\,\left|\frac{\partial\hat{x}(x^{(1)}, x^{(0)}, 
       \ldots , x^{(0)}; \ldots)}{\partial
       x^{12}}\right|^2\,{\varepsilon^{(0)}}^2 + k\,
     \left|\frac{\partial\hat{x}(x^{(1)}, x^{(0)}, \ldots , 
       x^{(0)}; \ldots)}{\partial
      x^{11}}\right|^2\,{\varepsilon^{(1)}}^2 } \nonumber \\
   {\varepsilon^{(1)}}' & = & \sqrt{kp}\,\left|\frac{\partial\hat{x}
     (x^{(0)}, \ldots , x^{(0)}; \ldots)}{\partial
      x^{11}}\right|\,{\varepsilon^{(0)}} \, .
\end{eqnarray} 
We note that the equations deliver precisely the spin-glass
instability of the liquid solution when we choose all local fields
equal to $x^*$. In the case of local fields $x^{(0)}$ and $x^{(1)}$ a
stability criterion can be derived by evaluating the eigenvalues of
the Jacobian matrix attributed to the iteration defined by
Eq. \eqref{gl_eps01it}. We find that for the inverse crystalline
solution $\varepsilon^{(0)} = \varepsilon^{(1)} = 0$ is a stable fixed
point of the iteration only if $\mu<\mu_{g, inv}\simeq 5.31$ which is
smaller than $\mu_g \simeq 5.89$. We can check this result via
population dynamics where we encode the distributions $P^{(0)}$ and
$P^{(1)}$ using two populations of fields. It turns out that the local
instability $\mu_{g, inv}$ is exact and furthermore we can show that
for $\mu<\mu_{g, inv}$ only the trivial field distributions
corresponding to the crystalline and inverse crystalline solutions
solve the 1RSB equations \eqref{gl_1RSBinvcr}. This is obviously in
contradiction with the results from section \ref{sec:compnumerics}
where the existence of a glassy crystalline phase is shown for
$\mu>\mu_{cg}\simeq 4.1$. We can thus conclude that the 1RSB formalism
is not sufficient to describe the onset of a glassy phase.

A way out of this dilemma might consist in passing to a
description in 2RSB which would mean to work with two distributions of
field distributions attributed to the sub-lattices. A 2RSB treatment
of the problem seems, however, complicated to handle. For the moment,
the construction of a ``glassy'' solution which would illuminate the
discrepancy between $\mu_{g, inv}$ and $\mu_{cg}$ remains an open problem.

\section{Close-packing limit ($\mu \to \infty$)}
\label{sec_closepack}

The 1RSB solution developed in the preceding section \ref{sec_cavity}
cannot be calculated analytically in the case of general $\mu$ where a
solution is obtained by a numerical method (population dynamics).

For the limiting case $\mu\to\infty$ which corresponds to a
close-packing, however, one can easily show that the transformed local
fields
\begin{equation} \label{gl_lokfeldneu}
  h^{jl} = \frac{1}{\mu}\,\ln(x^{jl}+1) =
  \frac{1}{\mu}\,\ln\left(\frac{\Xi_e^{jl}+\Xi_*^{jl}}{\Xi_e^{jl}}\right)
\end{equation}
must have a particularly simple distribution in the 1RSB formalism
which enables us to undertake an explicit calculation.

We first note that for $\mu\to\infty$ the grand-canonical partition
function is dominated by the terms that correspond to configurations
with a maximal number $N^{jl}_{e/*}$ of particles, {\em i.e.}
$\Xi^{jl}_{e/*} \propto e^{\mu N^{jl}_{e/*}}$ asymptotically with a
factor that counts the number of close-packings and does not depend on
$\mu$. This in mind, we find that Eq. \ref{gl_lokfeldneu} becomes
\begin{equation}\label{gl_neulflimit}
h^{jl} = \max (N^{jl}_e , N^{jl}_*) - N^{jl}_e 
\end{equation}
which implies that the local fields $h^{jl}$ must be nonnegative
integer numbers.

Using again the properties of the partition function for
$\mu\to\infty$ we can also rewrite the iteration equations
\ref{gl_rekzsumme} in terms of maximal numbers of particles:
\begin{eqnarray} \label{gl_anzahlrek0}
 N^{00}_e & = & \sum_{j=1}^{k} \max \Biggl( \sum_{l=1}^{p} N^{jl}_e , 
       \Bigl\{ N^{jl}_* + \sum_{\substack{m=1 \\ m \neq l}}^{p}
       N^{jm}_e \,\Big|\, l=1 , \ldots , p
        \Bigr\} \Biggr) \\ \label{gl_anzahlrek1}
 N^{00}_* & = & 1 + \sum_{j=1}^{k} \sum_{l=1}^{p} N^{jl}_e .
\end{eqnarray}
We use these relations to determine the maximal value of $h^{jl}$, putting
$j=0$ and $l=0$ without loss of generality. Following \eqref{gl_neulflimit} the
interesting case is $N^{00}_* > N^{00}_e$ where we find
\begin{eqnarray*}
h^{00} & = & N^{00}_* - N^{00}_e \\
       & = & 1 - \left( \sum_{j=1}^{k} \max \Biggl( \sum_{l=1}^{p} N^{jl}_e , 
       \Bigl\{ N^{jl}_* + \sum_{\substack{m=1 \\ m \neq l}}^{p}
       N^{jm}_e \,\Big|\, l=1 , \ldots , p
        \Bigr\} \Biggr) -  \sum_{j=1}^{k} \sum_{l=1}^{p}
      N^{jl}_e \right) \\
       & = & 1 - \sum_{j=1}^{k} \max \Bigl(0, \{N^{jl}_* - N^{jl}_e\,|\,
       l=1 , \ldots , p   \} \Bigr)
\end{eqnarray*}
which implies $h^{00}\le 1$. Thus we have shown that the transformed
local fields $h^{jl}$ satisfy $h^{jl} \in \{0 , 1\}$.

This result, which shows the interest of working with $h^{jl}$ instead
of $x^{jl}$, enables us to reduce drastically the degrees of freedom in
the 1RSB field distribution by making a simple one-parameter ansatz: 
\begin{equation} \label{gl_01ansatz}
P(h) = p_0\,\delta(h)+p_1\,\delta(h-1)
\end{equation}
where normalization requires $p_0 + p_1 = 1$.

The pair of parameters $p_0, p_1$ must be chosen such that the ansatz
solves the 1RSB cavity equation \eqref{gl_1RSBcavity} which can be
directly rewritten for the transformed local fields: 
\begin{equation} \label{gl_1RSBcavityneu}
P^{(m)}(h^{00}) \propto \int \prod_{j=1}^{k} \prod_{l=1}^{p} dh^{jl}
P^{(m)}(h^{jl})\,\delta\bigl(h^{00}-\hat{h}(h^{11} , \ldots , h^{kp})\bigr)
e^{-m\mu\Delta\Omega_{\text{iter}}(h^{11} , \ldots , h^{kp})} ,
\end{equation}
where the recursion relation $\hat{h}$ of the local fields and the
change in the grand-canonical potential $\Delta\Omega_{\text{iter}}$
per iteration step are now given by
\begin{equation}\label{gl_neulokfeldrek}
e^{\mu \hat{h}(h^{11} , \ldots , h^{kp})} = 1 +
\frac{e^{\mu}}{\prod_{j=1}^{k}\bigl(1-p+\sum_{l=1}^{p} e^{\mu h^{jl}}\bigr)}
\end{equation}
and
\begin{equation}\label{gl_neuomiter}
e^{-\mu\Delta\Omega_{\text{iter}}(h^{11} , \ldots , h^{kp})} =
\frac{\prod_{j=1}^{k}\bigl(1-p+\sum_{l=1}^{p} e^{\mu h^{jl}} \bigr) +
  e^{\mu}} {\prod_{j=1}^k \prod_{l=1}^p e^{\mu h^{jl}}}
\end{equation}
as follows directly from Eq. \eqref{gl_lokfeldrek} and
Eq. \eqref{gl_omiter}.

For practical reasons we define a vector $\vec{\nu}$ with $k$
components that encodes the relevant information of a local field
configuration $\{h^{11} , \ldots , h^{kp}\}$:

\begin{equation}
\nu_j \:\dot{=}\: \sum_{l=1}^{p} h^{jl} .
\end{equation}

As $h^{jl}$ satisfies $h^{jl}\in\{0,1\}$, $\nu_j$ counts the
non-vanishing fields in the $j$th clique. Furthermore we define a norm by
\begin{equation}
|\vec{\nu}| \: \dot{=} \: \#\{\nu_j \neq 0 \,|\, j = 1 , \ldots , k\} .
\end{equation}

We can see from Eq. \eqref{gl_neulokfeldrek} and
Eq. \eqref{gl_neuomiter} that $\hat{h}$ and
$\Delta\Omega_{\text{iter}}$ are functions of $\vec{\nu}$ only. We
calculate these quantities explicitly:
\begin{eqnarray}
\hat{h} & = & \frac{1}{\mu} \ln \left( 1 +
  \frac{e^{\mu}}{\prod_{j=1}^{k}\bigl(1-p+\sum_{l=1}^{p}e^{\mu h^{jl}}
    \bigr)}\right) \nonumber \\
        & = &  \frac{1}{\mu} \ln \left( 1 +
  \frac{e^{\mu}}{\prod_{j=1}^{k}\bigl(1-\nu_j+\nu_j e^{\mu}
    \bigr)}\right) \nonumber \\
        & \overset{\mu \to \infty}{=} & 
           \begin{cases}
              \;1 & \text{if } |\vec{\nu}| = 0 \\
              \;0 & \text{if } |\vec{\nu}| > 0
           \end{cases} \label{gl_inftyrek}
\end{eqnarray}
and
\begin{eqnarray}
\Delta\Omega_{\text{iter}} & = & \frac{1}{\mu} \ln \left(\frac
  {\prod_{j=1}^k \prod_{l=1}^p e^{\mu h^{jl}}} {\prod_{j=1}^{k}
    \bigl(1-p+\sum_{l=1}^{p} e^{\mu h^{jl}} \bigr) +
  e^{\mu}}\right) \nonumber \\
   & = & \sum_{j=1}^{k} \sum_{l=1}^{p} h^{jl} - 
  \frac{1}{\mu}\ln\left(\prod_{j=1}^{k}
  \Bigl(1-p+\sum_{l=1}^{p} e^{\mu h^{jl}} \Bigr) +
  e^{\mu} \right) \nonumber \\
   & = & \sum_{j=1}^{k} \nu_j - \frac{1}{\mu} \ln
   \left(\prod_{j=1}^{k}\bigl(1-\nu_j+\nu_j e^{\mu} \bigr) + e^{\mu}
   \right) \nonumber \\
   & \overset{\mu \to \infty}{=} &
        \begin{cases}
          \; -1  \quad \text{if } |\vec{\nu}| = 0 \\
          \; \begin{array}{l} 
                \sum_{j=1}^{k} \nu_j - |\vec{\nu}| \\
                = \sum_{j=1}^{k}\max (0, \nu_j-1) \\
                = \sum_{j=1}^{k}\max (1, \nu_j) - k 
                  \quad \text{if } |\vec{\nu}| > 0 
             \end{array} 
        \end{cases} \label{gl_inftyomit}
\end{eqnarray}
These results permit us to evaluate the 1RSB
Eq.~\eqref{gl_1RSBcavityneu} for the ansatz \eqref{gl_01ansatz} as
they reduce the integrations in the equation to a simple combinatoric
problem.

We see from Eq. \eqref{gl_inftyrek} that only the configuration
$\vec{\nu} = \vec{0}$ contributes to $h^{00} = 1$. With the ansatz for
$P(h)$ we deduce immediately from the 1RSB equation that
\begin{equation} \label{gl_contrp1}
p_1 \propto p_0^{k p} \, e^{y}
\end{equation} 
holds, where we have set $y \doteq m\mu$ and used
$\Delta\Omega_{\text{iter}}$ from Eq. \eqref{gl_inftyomit}.

In a second step we must sum up the contributions to $h^{00}=0$ that
come from all configurations with $\vec{\nu} \neq \vec{0}$. To this
purpose we note that there are 
\begin{equation*} 
 \prod_{j=1}^{k} {p \choose \nu_j}
\end{equation*}
different realizations of a configuration $\vec{\nu}$ with fields
$\{h^{11} , \ldots , h^{kp}\}$.

For each configuration we collect the probabilities emerging from the
ansatz yielding a factor 
\begin{equation*} 
 p_1^{\sum_{j=1}^{k} \nu_j} \, p_0^{k p - \sum_{j=1}^{k} \nu_j} .
\end{equation*}
In addition, we have to take into account the reweighting factor 
\begin{equation*}
  e^{-y \Delta\Omega_{\text{iter}}} = e^{y k}\, e^{-y \sum_{j=1}^{k}
    \max(1,\nu_j)} \, .
\end{equation*}

We obtain the contribution from a configuration $\vec{\nu} \neq
\vec{0}$ by taking the product of the three terms
\begin{equation*}
  \prod_{j=1}^{k} \left\{ {p \choose \nu_j} \, p_1^{\nu_j} \,
  p_0^{p-\nu_j} \, e^{y} \, e^{-y \max(1 , \nu_j)} \right\} \, .
\end{equation*}
The overall contribution to $h^{00}=0$ can now be calculated by
summation over the components of $\vec{\nu}\neq\vec{0}$:
\begin{eqnarray}
  p_0 & \propto & \sum_{\nu_1 = 0}^{p} \cdots \sum_{\nu_k = 0}^{p} 
  \:\prod_{j=1}^{k} \left\{ {p \choose \nu_j} \, p_1^{\nu_j} \,
  p_0^{p-\nu_j} \, e^{y} \, e^{-y \max(1 , \nu_j)} \right\}
  \underbrace{- \: p_0^{k p} }_{\substack{\text{correction for}\\ \nu_1
      = \cdots = \nu_k = 0}} \nonumber \\  
      & = & \prod_{j=1}^{k} \Biggl\{ \sum_{\nu_j=0}^{p} {p \choose
          \nu_j} \, p_1^{\nu_j} \,  p_0^{p-\nu_j} \, e^{y} \, e^{-y
          \max(1 , \nu_j)} \Biggr\} - p_0^{k p} \nonumber \\
      & = & \bigl\{ e^{y} (p_1 e^{-y} + p_0)^p \underbrace{- e^{y}
          p_0^p + p_0^p}_{\substack{\text{correction for} \\ \nu_j=0}}
          \bigr\}^k - p_0^{k p} . \label{gl_contrp0}
\end{eqnarray}
Making use of normalization  ($p_0+p_1=1$) we end up with the
iterative equations for the probabilities $p_0$ and $p_1$:
\begin{subequations}
\label{gl_rekpinfty}
\renewcommand{\theequation}{\theparentequation \roman{equation}}
\begin{eqnarray}
\hat{p}_0(p_0 , p_1) & = & \frac{ \bigl\{ e^{y} (p_1 e^{-y} + p_0)^p -
  p_0^p (e^{y}-1) \bigr\}^k - p_0^{k p}}{ \bigl\{ e^{y} (p_1 e^{-y} +
  p_0)^p - p_0^p (e^{y}-1) \bigr\}^k + p_0^{k p} (e^{y}-1) } \\
\hat{p}_1(p_0 , p_1) & = & \frac{p_0^{k p}\,e^{y}}{ 
  \bigl\{ e^{y} (p_1 e^{-y} +
  p_0)^p - p_0^p (e^{y}-1) \bigr\}^k + p_0^{k p} (e^{y}-1) } .
\end{eqnarray}
\end{subequations}
The equations can be solved numerically for given $y$ with the help of
a computer algebra system.

It remains the question which value $y = y_s$ we must choose in order
to describe thermodynamic equilibrium. The answer is given in section
\ref{sec_cavity} via the condition that the configurational entropy
must vanish, {\em i.e.} $\Sigma(y_s) \overset{!}{=} 0$. We use
Eq. \ref{gl_partialmphi} with the substitution $y = m\mu$ to find
\begin{equation}
 \Sigma(y) = y^2 \partial_y\Phi(y) \,. \label{gl_param1cp}
\end{equation}
which provides $\partial_y\Phi(y = y_s) = 0$ as an alternative
condition for $y_s$.

Consequently, we shall compute $\Phi(y)$ to complete the picture of
the close-packing limit. $\Phi(y)$ is given by
Eq. \eqref{gl_bethepeiphi} with Eq. \eqref{gl_1RSBdpsite} and
\eqref{gl_1RSBdpsite} as well as Eq. \eqref{gl_appdelomallg}. We
proceed analogously to the derivation of the equations for $p_0, p_1$:
in a first step, we evaluate Eq. \eqref{gl_appdelomallg} with
transformed local fields in the limit $\mu\to\infty$. In a second
step, the integrations in Eq. \eqref{gl_1RSBdpsite} and
\eqref{gl_1RSBdpsite} are performed. This reads:
\begin{eqnarray}
  \Delta\Omega_{\text{link}} & = & \sum_{l=1}^{p+1} h^{(l)} -
  \frac{1}{\mu} \ln \left(\sum_{l=1}^{p+1}e^{\mu h^{(l)}} - p \right)
  \nonumber \\
  & = & \nu - \frac{1}{\mu} \ln (1-\nu+\nu e^{\mu}) \nonumber \\
  & \overset{\mu \to \infty}{=} & 
     \begin{cases}
       \:0 & \text{if } \nu = 0 \\
       \:\nu - 1 & \text{if } \nu > 0
     \end{cases} \: ,
\end{eqnarray}
where we have set $\nu\:\dot{=}\: \sum_{l=1}^{p+1} h^{(l)}$. Further
on we have
\begin{eqnarray}
  \Delta\Omega_{\text{site}} & = & \sum_{j=1}^{k+1} \sum_{l=1}^{p}
  h^{jl} - \frac{1}{\mu} \ln \left( \prod_{j=1}^{k+1}\biggl(1 - p +
      \sum_{l=1}^{p} e^{\mu h^{jl}}\biggr) + e^{\mu} \right) \nonumber
    \\
  & = & \sum_{j=1}^{k+1} \nu_j - \frac{1}{\mu} \ln
  \left(\prod_{j=1}^{k+1} \bigl(1-\nu_j+\nu_j e^{\mu}\bigr) 
  + e^{\mu} \right)
  \nonumber \\
  & \overset{\mu \to \infty}{=} & 
     \begin{cases}
       \: -1 & \text{if } |\vec{\nu}| = 0 \\
       \: \sum_{j=1}^{k+1} \max(1, \nu_j) - (k+1) & \text{if }
       |\vec{\nu}| > 0
     \end{cases} \: .
\end{eqnarray}
Note that the expression for $\Delta\Omega_{\text{site}}$ is obtained
from $\Delta\Omega_{\text{iter}}$ by the substitution  $k
\rightarrow k+1$. We perform the integrations to obtain
\begin{eqnarray}
  e^{-y \Delta\Phi_{\text{link}}} & = & p_0^{p+1}\,e^{-y \cdot 0} +
  \sum_{\nu=1}^{p+1} {p+1 \choose \nu}\,
  p_1^{\nu}\,p_0^{p+1-\nu}\,e^{-y\cdot(\nu-1)} \nonumber \\
    & = & e^y\, (p_1 e^{-y}+p_0)^{p+1} - p_0^{p+1} (e^y-1) \\
\nonumber \\
\text{and} \quad e^{-y \Delta\Phi_{\text{site}}} & = & \bigl\{ e^{y}
(p_1 e^{-y} + p_0)^p - p_0^p (e^{y}-1) \bigr\}^{k+1} + p_0^{(k+1) p}
(e^{y}-1) \, , 
\end{eqnarray}
where the latter term also results from the sum of the contributions
\eqref{gl_contrp1} and \eqref{gl_contrp0} after substituting $k
\rightarrow k+1$.

The knowledge of $\Phi(y)$ can serve to access the particle density $\rho$: 
\begin{equation} 
 \rho = \partial_{\mu} \ln \Xi = -\partial_{\mu}(\mu\Phi) =
 -\partial_y(y \Phi) . \label{gl_param2cp}
\end{equation}
Specially in thermodynamic equilibrium the particle density is given by
\begin{equation}
  \rho_s = \rho(y_s) = -\Phi(y_s) ,
\end{equation}
as $\partial_y\Phi$ vanishes at $y_s$.

\begin{figure}[htbp]
  \begin{center}
    \includegraphics[height = 8cm]{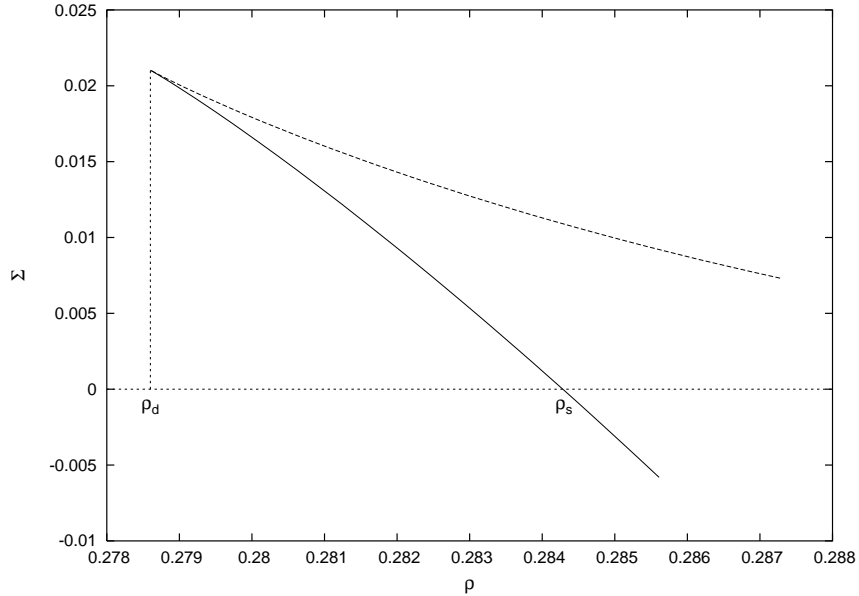}
    \caption{1RSB solution for the complexity $\Sigma$ as a function
      of the particle density $\rho$ in the limit $\mu \to \infty$
      ($k=3, p=2$). $\Sigma$ being a concave function, only the full
      line has physical meaning. $\Sigma<0$ is attributed to untypical
      graphs that have vanishing statistical weight.}
\label{fig_complex}
  \end{center}
\end{figure}

Eq. \eqref{gl_param2cp} together with Eq. \eqref{gl_param1cp} deliver
the complexity $\Sigma(\rho)$ as a curve parameterized by $y$
(Fig. \ref{fig_complex}). The complexity $\Sigma$ has a maximum at
$\rho_d<\rho_s$ while it vanishes for the equilibrium density
$\rho_s$. The meaning of $\rho_d$ for the dynamics of the system, in
particular for dynamics induced by local algorithms, is discussed in
the subsequent section.

Some results for the particle density of a close-packing
$\rho_{\infty} \equiv \rho_s$ and the density $\rho_d$ with maximal
complexity as obtained in the 1RSB solution are shown for different
$k$ and $p$ in Fig. \ref{fig_closep}.

\subsection{Comparison with numerical bounds}
\label{sec:numericCompact}

\begin{figure}[htb]
  \begin{center}
    \includegraphics[height = 8cm]{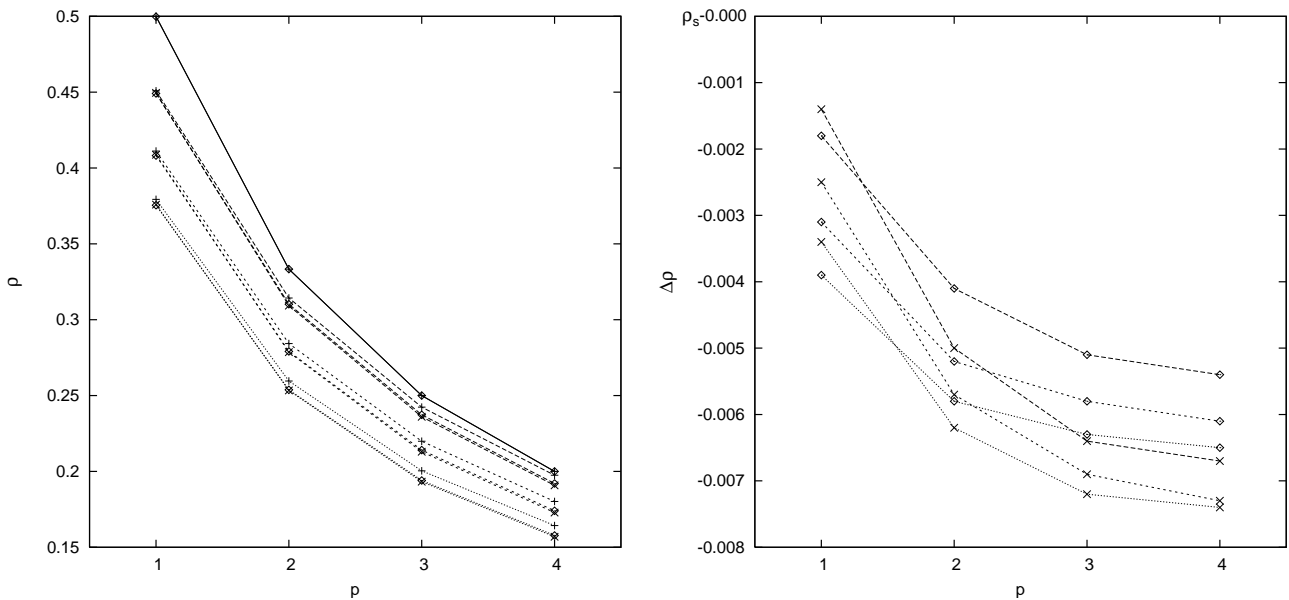}
    \caption{Left: particle density $\rho_{\infty}$ of a close-packing ($+$)
      and density $\rho_d$ with maximal complexity ($\times$) from the
      1RSB solution compared with numerical lower bounds ($\Diamond$)
      for $\rho_{\infty}$ (from top to bottom: $k = 1, 2, 3, 4$). For
      $k=1$, $\rho_{\infty}$ and $\rho_d$ coincide. Right: the
      difference of $\rho_d$ ($\times$) and the numerical bound
      ($\Diamond$) to
      $\rho_{\infty}\equiv\rho_s$ is shown for $k = 2, 3, 4$ (top to bottom).}
\label{fig_closep}
  \end{center}
\end{figure}

We have compared the analytical 1RSB results for $\rho_{\infty}$ and
$\rho_d$ with numerical lower bounds for $\rho_{\infty}$ that were
obtained in Monte-Carlo simulations. For given $k$ and $p$ random
graphs of size $N={(p+1)} \cdot 500\,000$ that have locally the
structure of a generalized Bethe lattice were generated. Roughly
speaking, this can be achieved by iterating the following procedure
until every vertex has $(k+1)p$ neighbors: choose randomly $p+1$
vertices with less then $(k+1)p$ neighbors and fully connect them to
form a clique. Furthermore we introduce {\em local} dynamics on the
lattice by applying the following local rules to a randomly chosen
site: (i) if the site is empty: introduce a particle, (ii) if the site
is occupied: move the particle to a randomly chosen neighboring
site. If the application of a rule conflicts with the hard-core
constraint, the rule is not applied. The time $1\,\text{MCs}$
corresponds again to choosing $N$ times a site at
random. Fig.~\ref{fig_closep} shows the particle density that the
initially empty system ($t = 0\,\text{MCs}$) has reached at MC-time $t
= 100\,000\,\text{MCs}$. The error can be estimated to less than
$0.1\,\%$ by comparing the results for different realizations of the
graph. The asymptotic behavior of the density as a function of time
can be characterized by the following observations. The density at $t
= 50\,000\,\text{MCs}$ ($t = 75\,000\,\text{MCs}$) is about
$99.96\,\%$ ($99.99\,\%$) of the density at $t = 100\,000\,\text{MCs}$
for small $p$ and about $99.86\,\%$ ($99.95\,\%$) for great $p$. These
figures illustrate that the compaction process becomes very slow in
the second half of the simulated time frame. We do not expect the
system to equilibrate on a finite time scale due the presence of
glassy states. The quality of the numerical results, which constitute
lower bounds for the exact values of $\rho_{\infty}$, is thus
determined by the finiteness of the simulated systems and the
simulated times which were chosen maximal within the limits of
reasonable computation time.

We notice that the 1RSB results for
$\rho_{\infty}$ are never violated by the numerical bounds. In
the case $k=1$ the agreement between numerical and analytic results is
very good which can be understood {\em e.g.}  assuming $p=1$ where the
``random'' graph forms simply a circle which can be always packed in
the crystalline ground state up to a misfit if $N$ is odd. For $k>1$
the numerical bound stays always below $\rho_{\infty}$ with a gap that
grows with increasing $p$ up to $4.0\,\%$. On the other hand, we find
a good agreement of the numerical bounds with the results for
$\rho_d$. We shall explain this observation with a remark on the
configuration space structure as obtained within the 1RSB solution.

We have assumed as a starting point for the derivation of the 1RSB
cavity equation that a non-vanishing complexity is equivalent to the
presence of an exponential (in $N$) number of metastable states in the
configuration space ({\em ergodicity breaking}). As different
metastable states are only connected via configurations of lower
densities any local algorithm, trying to increase the number of
particles in every step, gets trapped in a metastable state. Using the
definition of the complexity \eqref{eq:complexity} it becomes clear
that the state in question will have almost always a particle density
$\rho\simeq\rho_d<\rho_{\infty}$ which maximizes the complexity
(cf. Fig. \ref{fig_complex}). The numerical bounds that result from a
local algorithm give indeed strong indication that the concepts of the
1RSB formalism have a very concrete meaning for algorithmically
induced as well as physical dynamics.

Note that there is recent work by Montanari and Ricci-Tersenghi who
performed numerical cooling experiments for $p$-spin glasses
\cite{montaRicci04}, a model which is somewhat similar to our lattice
gas model. The behavior for different cooling-rates underlines the
relevance of the so called iso-complexity curves. In this context, the
numerical experiments performed in our study constitute a limiting
case of very high compaction-rate which would lead to densities
significantly below $\rho_{\infty}$ if the particle movement, which
has no equivalent in the spin-glass context, was omitted. An
investigation for different compaction-rates might give interesting
complementary insights concerning the glassy phase of the lattice gas.

We conclude with the remark that there is {\em a priori} no indication
for the correctness of the results that were calculated under the
restrictions of the 1RSB ansatz. In the next section we shall sketch
the baselines of a stability theory for 1RSB in order to check the
reliability of the preceding results.

\subsection{Stability analysis of the 1RSB solution}
\label{sec_stabi1RSB}

In this section we give a brief summary of concepts concerning
the local stability analysis of a 1RSB solution towards a 2RSB
solution. A detailed discussion of the topic was given by Rivoire
{\em et al.} \cite{rivoire} for the case a Bethe lattice gas which
shares common properties with the system under study.

We start with the introduction of the 2RSB formalism. Compared to the
1RSB solution which was based on the organization of configurations in
states, 2RSB acts on the level of states by assembling states in
clusters to which we attribute grand-canonical potentials $\omega_c$. The
number of states within a cluster is given by referring to $\omega_c$:
\begin{equation}
  \mathcal{N}_s(\omega_s) \sim e^{Ny_s(\omega_s-\omega_c)} ,
\end{equation}
where $y_s$ must be understood in analogy to Eq. \eqref{gl_mom0def} as
the slope of the complexity of the states within the cluster. Using
this relation as a starting point, the 1RSB cavity equation can be
derived with the same arguments as in section \ref{sec_cavity} yielding
\begin{equation}
  \hat{P}[\{P^{jl}\}](x) \propto \int \prod_{j=1}^{k} \prod_{l=1}^{p} dx^{jl}
  P^{jl}(\{x^{jl}\}) \delta(x - \hat{x}(\{x^{jl}\})) \,
  e^{-y_s\Delta\Omega_{\text{iter}}(\{x^{jl}\})} \, ,
\end{equation}
when we discard the homogeneity assumption of the distributions
$P^{jl}$. The shape of the equation invites to iterate the transfer
which starts with the recursion $\hat{x}(\{x^{jl}\})$ to deliver the
1RSB equation through the introduction of field
distributions. Concretely, this means to start with the recursion
$\hat{P}[\{P^{jl}\}]$ and assume distributions of field distributions
(clustering). As in the 1RSB derivation we introduce the number of
clusters with grand-canonical potential $\omega_c$ with respect to a
reference point $\omega_0$:
\begin{equation}
  \mathcal{N}_c(\omega_c) \sim e^{Ny_c(\omega_c-\omega_0)}
\end{equation}

The clustering is realized by the introduction of distributions $Q[P]$
for the field distributions. Using the arguments from section
\ref{sec_cavity} the so-called 2RSB cavity equations can be derived:  
\begin{equation}
 Q[P] \propto \int \prod_{j=1}^{k} \prod_{l=1}^{p} \mathcal{D}\!P^{jl}\,
 Q[P^{jl}] \delta(P-\hat{P}[\{P^{jl}\}]) \,
 e^{-y_c \Delta\Phi_{\text{iter}}[\{P^{jl}\}]} \: .
\end{equation}

where $\Delta\Phi_{\text{iter}}$ is the change in grand-canonical
potential corresponding to one step of the iteration
$\hat{P}[\{P^{jl}\}]$. It can be obtained by evaluating
$\Delta\Omega_{\text{iter}}$ with respect to the distributions $P^{jl}$:
\begin{equation}
  e^{-y_s \Delta\Phi_{\text{iter}}[\{P^{jl}\}]} = \int \prod_{j=1}^{k}
  \prod_{l=1}^{p} dx^{jl} P^{jl}(x^{jl}) e^{-y_s
    \Delta\Omega_{\text{iter}}(\{x^{jl}\})} .
\end{equation} 

The hierarchical scheme of distributions can be continued to higher
levels of RSB, leading to full replica symmetry breaking ({\em frsb})
in the limit of infinite continuation. A proof for the correctness of
a 1RSB solution would require to show that every description in higher
levels of RSB coincides with the 1RSB description. This is of course
not feasible if only the 1RSB solution is explicitly known. In this
case one can show the local stability of the 1RSB solution toward
2RSB which is believed to be a good indication for the correctness of
the 1RSB solution as no example for a discontinuous transition from
1RSB to RSB of higher levels is known so far.

It was shown be Montanari and Ricci-Tersenghi that there are two
possible kinds of instabilities that may occur
\cite{montaRicci}. These result from two different descriptions of a
1RSB solution within the 2RSB formalism. One way is to assume the
existence of a single trivial cluster where the states are given by
$P^*(x)$, with $P^*$ being the 1RSB solution. Thus we have $Q[P]=
\delta[P-P^*]$ and the corresponding instability can be detected by
making the ansatz $Q[P] = f[P-P^*]$ where $f$ is a functional with
support around the null function. This corresponds to an assembling of
the states that organize in clusters. The process is called {\em
aggregation of states} and the instability is classified type I. The
second possibility consists in encoding the field distribution
$P^*(x)$ in the cluster structure while the states are trivial. For
discrete $P^*$ we write $P^* = \sum_a p_a\delta_a$ with
$\delta_a(\cdot) = \delta(\cdot - a)$. The cluster structure is then
given by $Q[P] = \sum_a p_a \delta[P-\delta_a]$ and the instability is
detected by $Q[P] = \sum_a p_a f_a[P-\delta_a]$ with $f_a$ having
support around the null function. At the instability each state can be
considered as a germ giving birth to a cluster. The process is called
{\em fragmentation of states} and the instability is classified type
II.

We shall not stress on the details of the concrete calculations which
can be found in the literature for various models including
combinatorial problems 
\cite{rivoire,montaRicci,duboisMonasson,mezardParZecc,montaParRicci,mertens,pagnani}.

In order to investigate the stability of the 1RSB solution in the
close-packing limit ($\mu\to\infty$) we focus on the type-I
instability. We note that there is a strong analogy between the
calculation of the spin-glass instability, which was detected by
putting $P(x) = f(x-x^*)$ where $f$ is a smeared $\delta$-function,
and the ansatz $Q[P] = f[P-P^*]$ suitable to find the type I
instability. By transporting the calculations for the spin-glass
instability to the next higher level in the RSB hierarchy the
stability criterion  $k p\,|\lambda_{\text{max}}(y)|^2 < 1$, which reads precisely like
Eq. \eqref{gl_Spinglasinst}, has been derived by Rivoire
{\em et al.} \cite{rivoire}. Here $\lambda_{\text{max}}(y)$ is the
eigenvalue of greatest modulus of the matrix
\begin{equation}
  A(y) = \left (
         \begin{array}{cc}
           \frac{\partial \hat{p}_0}{\partial p_0} &  
           \frac{\partial \hat{p}_0}{\partial p_1} \\ \vspace{-0.3cm}\\
           \frac{\partial \hat{p}_1}{\partial p_0} &  
           \frac{\partial \hat{p}_1}{\partial p_1}
         \end{array}
      \right ) \; , \nonumber
\end{equation}
which is the Jacobian matrix associated with the iteration of the 1RSB
field distributions Eq. \eqref{gl_rekpinfty}.

A computation of $\lambda_{\text{max}}(y)$ yields that we have $k p\,
|\lambda_{\text{max}}(y)|^2 > 1$ at $y = y_d$ as well as $y = y_s$ for
all the cases shown in Fig. \ref{fig_closep}, {\em i.e.} the 1RSB
solution is locally unstable toward 2RSB. We can conclude that the
results for $\rho_d$ and $\rho_{\infty}$ that were derived in the 1RSB
context are not exact. Following the general belief that the
instability of a 1RSB solution indicates {\em frsb}, this would even
mean $\rho_s = \rho_d$ for $\mu\to\infty$, which is not suggested by
the numerical findings. It must however be considered that a system
cannot be brought instantaneously to infinite $\mu$ via numerical
compaction. Under the action of a MC-algorithm the simulated system
will pass through a chain of genuine non-equilibrium states. By
mapping these on equilibrium states according to their density we can
approximately attribute a chemical potential $\mu(t)$ to the
system. It might seem more adequate to rely on a stability analysis of
the 1RSB solution for $\mu=\mu_d$ in order to decide if the compacted
system is governed by metastable states. Unfortunately, this analysis
seems to be out of reach for present analytical techniques.

\section{Conclusion and outlook}

To conclude, we have investigated in great detail the thermodynamic
behavior of a hard-sphere lattice gas model on generalized Bethe
lattices.

On usual Bethe lattices, which do not contain (short) loops, the model
shows a very simple behavior: At low density, the system is found in a
liquid phase, and the local density is homogeneous. At a certain
point, the system undergoes a second-order freezing transition. The
character of this transition depends on the global graph structure:
The frozen phase is either crystalline (if long loops of odd length
are almost absent in the graph), or it is given by a spin-glass phase
(if the Bethe lattice is defined as a regular random graph, and the
crystalline structure is inconsistent with the large-scale structure
of the graph).

The introduction of short loops changes this behavior drastically. We
have therefore studied generalized Bethe lattices, which are locally
characterized by the existence of many short loops, but on a
coarse-grained level form locally tree-like hyper-graphs. This mixed
structure allows for an analytical solution of the hard-sphere lattice
gas.

Again, the low-density phase describes a spatially homogeneous liquid.
If compactified, the system undergoes either a first-order
crystallization transition or a discontinuous glass transition which
is characterized by a broken replica symmetry, and by the existence of
an exponentially large number of metastable states. We also find more
exotic high-density phases which are at least metastable and can thus
be found also in numerical simulations. The first of these phases is
an inverse-crystalline one. Its sublattice densities are inverted, as
compared to the densest crystalline packing. For even higher density,
this phase becomes unstable with respect to a crystalline glass
transition: Even if the local densities are frozen to random values,
their distributions are organized in a regular, i.e. inverse
crystalline manner. 

This means that hard-sphere lattice gases on generalized Bethe
lattices may serve as valid, microscopically motivated and
analytically tractable mean-field models for the glass transition. Do
they, however, also present good approximations to the behavior in low
dimensions? The hard-sphere model is in fact well studied on various
two- and three-dimensional lattices, the most famous cases are the
hard-hexagon model and its solution by the corner-transfer matrix
\cite{Bax1,Bax2,Bax3}, and the three-dimensional cubic lattice
\cite{Gaunt} which can be analyzed by means of series
expansions. These two cases show second-order crystallization
transitions. In \cite{weigt} we have, however, introduced a
finite-dimensional lattice mimicking the structure of the generalized
Bethe lattice, and this model shows in fact a glass-like slowing
down. The latter is, however, much less pronounced due to the
existence of activated processes in finite dimensions.

At this point, another remark concerning the relation of the
generalized Bethe-lattice models to $p$-spin glass models is
necessary. The latter model class was extensively used in particular
in the last decade in the context of the glass transition
\cite{BoCuKuMe,Cu}. On one hand, they allow for an analytical
treatment of the model statics and dynamics, the latter being formally
equivalent to the mode-coupling approach to more realistic
glass-forming liquids. $p$-spin glasses are, however, completetly
missing a microscopic motivation. One point here is the existence of
unphysical multi-spin interactions. If we look, however, to our
generalized Bethe-lattice model, the introduction of cliques
effectively introduces multi-site interactions -- only one site per
clique can be occupied -- even if these are defined via groups of pair
interactions. This allows to reinterprete also the $p$-spin
interactions in spin-glass models as a kind of coarse grained
effective interaction.

All the considerations in this paper are restricted to the
thermodynamic, i.e.~equilibrium behavior of the model. In particular
in the glassy regions, we know that the dynamics becomes very slow,
and the system is practically always out of equilibrium. The
analytical description of the dynamics of finite-connectivity models
is, however, a much more involved problem, and it is to a large extent
an unsolved one. In a subsequent publication, we will confront our
findings on the static behavior with an approximate analysis of the
dynamics \cite{self}. The approach taken there is related to a
projective approximation scheme introduced in \cite{SeMo,BaHaWe} in
the context of algorithms for combinatorial optimization, and then
applied to simple Ising ferromagnets in \cite{SeWe}. Within this
scheme, the dynamics is projected to the dynamics of a small set of
global observables, and the resulting equations are closed on the
basis of a pseudo-equilibrium ansatz in an enlarged ensemble. The
selection of this generalized ensemble, and the technical realization
of the projection are based on the tools developed in this article,
and therefore will form a natural continuation of the present work.

{\bf Acknowledgment:} We are grateful to A.K. Hartmann, G. Semerjian
and A. Zippelius for many interesting discussions. HHG acknowledges
also the hospitality of the ISI Foundation in Turin, where some of the
final steps in this project were done.


\begin{thebibliography}{1}
 
\bibitem{Go} W. G\"otze, {\it Liquids, freezing, and glass
  transition}, Les Houches (1989).

\bibitem{An} C.A. Angell, Science {\bf 267}, 1924 (1995).

\bibitem{BoCuKuMe} J.P. Bouchaud, L.F. Cugliandolo, J. Kurchan, and
  M. M\'ezard, in {\it Spin Glasses and Random Fields}, ed. A.P. Young
  (World Scientific, Singapore, 1998).

\bibitem{RiSo} F. Ritort and P. Sollich, Adv. Phys. {\bf 52}, 219
  (2003).

\bibitem{Cu} L.F. Cugliandolo, {\it Dynamics of glassy systems},
  Lecture notes, Les Houches (2002).

\bibitem{biroli}
G.~Biroli and M.~M{\'e}zard,
\newblock Phys. Rev. Lett. \textbf{88}, 025501 (2002).
                                                                                
\bibitem{weigt}
M.~Weigt and A.~K.~Hartmann,
\newblock Europhys. Lett. {\bf 62}, 533 (2003).
                                                                                
\bibitem{pica}
M.~Pica~Ciamarra, M.~Tarzia, A.~de~Candia ans A.~Coniglio,
\newblock Phys. Rev. E \textbf{67}, 057105 (2003).
                                                                                
\bibitem{rivoire}
O.~Rivoire, G.~Biroli, O.~C.~Martin and M.~M{\'e}zard,
\newblock Eur. Phys. J. B \textbf{37}, 55-78 (2004).


\bibitem{SeBiTo} M. Sellitto, G. Biroli, and C. Toninelli, to app. in
  Europhys. Lett. (2005).
 
\bibitem{mezparvir}
M.~M{\'e}zard, G.~Parisi and M.~A.~Virasoro,
\newblock \textit{Spin Glass Theory and Beyond},
\newblock (World Scientific, Singapore, 1987).

\bibitem{mezpar}
M.~M{\'e}zard and G.~Parisi,
\newblock Eur. Phys. J. B {\bf 20}, 217-233 (2001).

\bibitem{self} H.~Hansen-Goos and M.~Weigt, in preparation.

\bibitem{runnels}
L.~K.~Runnels,
\newblock J. Math. Phys. {\bf 8}, 2081 (1967).


  
\bibitem{monasson}
R.~Monasson,
\newblock Phys. Rev. Lett. \textbf{75}, 2847-2850 (1995).
 
\bibitem{montaRicci04} A.~Montanari and F.~Ricci-Tersenghi,
Phys. Rev. B {\bf 70}, 134406 (2004).
                                                                      
\bibitem{duboisMonasson}
O.~Dubois, R.~Monasson, B.~Selman and R.~Zecchina (eds.),
\newblock Theor. Comp. Sci. {\bf 265}, issue 1-2 (2001).

                                                                               
\bibitem{mezardParZecc}
M.~M{\'e}zard, G.~Parisi and R.~Zecchina,
\newblock Science \textbf{297}, 812 (2002);
\newblock M.~M{\'e}zard and R.~Zecchina,
\newblock Phys. Rev. E \textbf{66}, 056126 (2002).

                                                                                
   
\bibitem{montaRicci}
A.~Montanari and F.~Ricci-Tersenghi,
\newblock Eur. Phys. J. B \textbf{33}, 339 (2003).
         
\bibitem{montaParRicci}
A.~Montanari, G.~Parisi and F.~Ricci-Tersenghi,
\newblock J. Phys. A: Math. Gen. \textbf{37}, 2073-2091 (2004).
                                                                                
\bibitem{mertens}
S.~Mertens, M.~M{\'e}zard and R.~Zecchina,
\newblock preprint \texttt{cs.CC/0309020} (2003).
                                               
\bibitem{pagnani}
F.~Krz\c{a}ka{\l}a, A.~Pagnani and M.~Weigt,
\newblock Phys. Rev. E {\bf 70}, 046705 (2004).



\bibitem{Bax1} R.J. Baxter, J. Phys. A {\bf 13}, L61 (1980).

\bibitem{Bax2} R.J. Baxter, {\it Exactly Solved Models in Statistical
Mechanics}, (Academic Press, London 1982).

\bibitem{Bax3} R.J. Baxter, Ann. Comb. {\bf 3}, 191 (1999).

\bibitem{Gaunt} D.S. Gaunt, J. Chem. Phys. {\bf 46}, 3237 (1967).

\bibitem{SeMo} G. Semerjian and R. Monasson, Phys. Rev. E {\bf 67},
066103 (2003).

\bibitem{BaHaWe} W. Barthel, A.K. Hartmann, and M. Weigt, Phys. Rev. E
{\bf 67}, 066104 (2003).

\bibitem{SeWe} G. Semerjian and M. Weigt, J. Phys. A {\bf 37}, 5525
(2004).

\end{thebibliography}
\end{document}